\documentclass[journal]{IEEEtran}

\usepackage[nocompress]{cite}

\usepackage{amsmath,amssymb,amsfonts}
\usepackage{algorithm}
\usepackage{algpseudocode}

\usepackage{textcomp}
\usepackage[normalem]{ulem}

\usepackage{graphicx} 
\usepackage{mathtools}

\usepackage{xcolor}

\usepackage{tikz}
\usepackage{hyperref}

\usepackage{booktabs}
\usepackage{tabularx}
\usepackage{adjustbox}
\usepackage{subcaption}
\usetikzlibrary{matrix}
\usepackage{comment}
\usepackage{soul}

\def\BibTeX{{\rm B\kern-.05em{\sc i\kern-.025em b}\kern-.08em
    T\kern-.1667em\lower.7ex\hbox{E}\kern-.125emX}}
\begin{document}

\title{An Efficient Outlier Detection Algorithm for Data Streaming \\
}

\author{Rui~Hu,
        Luc~(Zhilu)~Chen,
        Yiwei~Wang
\thanks{R. Hu is with California State University, San Bernardino, San Bernardino, CA, USA. Email: rui.hu@csusb.edu}
\thanks{L. Chen is with John A. Paulson School of Engineering and Applied Sciences, Harvard University, Cambridge, MA, USA. Email: lucchen@g.harvard.edu} 
\thanks{Y. Wang is with University of California, Riverside, Riverside, CA, USA. Email: yiweiw@ucr.edu }
}

\maketitle

\begin{abstract}
The nature of modern data is increasingly real-time, making outlier detection crucial in any data-related field, such as finance for fraud detection and healthcare for monitoring patient vitals. Traditional outlier detection methods, such as the Local Outlier Factor (LOF) algorithm, struggle with real-time data due to the need for extensive recalculations with each new data point, limiting their application in real-time environments. While the Incremental LOF (ILOF) algorithm has been developed to tackle the challenges of online anomaly detection, it remains computationally expensive when processing large streams of data points, and its detection performance may degrade after a certain threshold of points have streamed in. In this paper, we propose a novel approach to enhance the efficiency of LOF algorithms for online anomaly detection, named the Efficient Incremental LOF (EILOF) algorithm. The EILOF algorithm only computes the LOF scores of new points without altering the LOF scores of existing data points. Although exact LOF scores have not yet been computed for the existing points in the new algorithm, datasets often contain noise, and minor deviations in LOF score calculations do not necessarily degrade detection performance. In fact, such deviations can sometimes enhance outlier detection. We systematically tested this approach on both simulated and real-world datasets, demonstrating that EILOF outperforms ILOF as the volume of streaming data increases across various scenarios. The EILOF algorithm not only significantly reduces computational costs, but also systematically improves detection accuracy when the number of additional points increases compared to the ILOF algorithm.
\end{abstract}


\section{Introduction}
Outlier detection is a crucial process in data analysis, aimed at identifying data points that significantly deviate from the majority, often indicating abnormal behavior, errors, or novel discoveries \cite{hawkins1980identification, panjei2022survey, sejr2021explainable}. It serves as a fundamental step not only in data preprocessing to ensure data quality and integrity but also as a pivotal component in various machine learning tasks to improve model accuracy and robustness. Outlier detection techniques are broadly classified into model-based, distance-based, density-based, and subspace-based methods\cite{boukerche2020outlier, xu2018comparison, tang2017local, marques2023evaluation}. 

As a density-based method, the Local Outlier Factor (LOF) algorithm \cite{breunig2000lof} is widely esteemed for its ability to identifies outliers by comparing a point’s local density to that of its neighbors. The key step in the LOF algorithm is to calculate a LOF score for each point, which measures the difference between the density estimate for a point $p$ in a database $D$ and the average density estimates for the $k$-nearest neighbors of $p$ \cite{schubert2014generalized}.
The local density estimate is constructed based on reachability distance between two points (see Eq. (\ref{reach-dis}) for definition). This method is adept at uncovering local anomalies that may not be detected by global threshold-based approaches \cite{breunig2000lof}. LOF and similar algorithms have been extensively applied across various fields, including fraud detection, network security, and health diagnostics, demonstrating their versatility and importance in both data preprocessing and in enhancing machine learning models by removing or flagging anomalous data that could lead to skewed predictions \cite{chandola2009anomaly, cao2021scalable}.

However, LOF-based methods face challenges in online settings, where data continuously streams in, requiring repetitive recalculations of LOF scores when new data points arrive \cite{alsini2021grid, reunanen2020unsupervised, sadik2014research}. 
To address this challenge, the Incremental Local Outlier Factor (ILOF) algorithm was proposed in \cite{pokrajac2007incremental}. The idea of ILOF is to update the reachability distances of all points whose k-nearest neighbors include the new data point, ensuring that the local density estimates reflect the latest changes. 
As a result, the algorithm achieves equivalent detection performance to the iterated static LOF algorithm,
which re-applies the LOF algorithm after insertion of new data points, but with significantly reduced computational time. However, the computational cost of ILOF might still be large, particularly when a significant amount of points stream in. For instance, it may require a large amount of memory to keep all the previous points \cite{salehi2016fast}. Several enhancements to ILOF have been proposed, such as the Memory Efficient Incremental Local Outlier (MiLOF) detection algorithm, which manages data streams within a fixed memory bound while maintaining accuracy close to ILOF but with better memory and time efficiency.

Numerical experiments show that for the ILOF algorithm, detection accuracy may decrease as the volume of streaming data grows, which indicates that, in data streaming settings, recalculating precise LOF scores for the entire dataset may not always improve detection accuracy. Motivated by this observation, we propose an efficient and robust approach to the Local Outlier Factor (LOF) algorithm for data streaming, named the Efficient Incremental Local Outlier Factor (EILOF) algorithm. Unlike the ILOF algorithm, which essentially (re)-calculates LOF scores for all points whose k-nearest neighbors include the new data points, our algorithm only computes the LOF scores of new points without altering the LOF scores of existing points. The approach is inspired by methods that focus on improving computational efficiency in data stream analysis while preserving detection accuracy \cite{gama2010knowledge, domingos2000mining, cacciarelli2024active}. This computational strategy gains significant efficiency, particularly when more data streams in. Although we may not obtain accurate LOF scores for all points in the updated dataset, minor deviations in LOF score computations do not necessarily lead to a degradation in detection performance, as datasets inherently contain noise. In practice, as shown by our numerical experiments, detection performance may improve without accurately updating the LOF scores of existing data points. One particular reason is that when more data streams in, the original $k$, the number of nearest neighbor, is no longer optimal for the new dataset.

Enhanced Incremental Local Outlier Factor (EILOF) addresses the challenges of streaming data by processing new observations incrementally, updating outlier scores and neighborhood structures without requiring full dataset recalculations. This approach significantly reduces computational overhead, enabling near real-time anomaly detection for high-frequency data streams. Additionally, EILOF enhances scalability through the use of optimized data structures and approximation methods, ensuring robust performance in dynamic environments. These features make EILOF particularly effective for applications such as fraud detection, sensor monitoring, and network intrusion detection.

The primary contributions of this work are as follows:
\begin{itemize}
    \item Proposing the Efficient Incremental Local Outlier Factor (EILOF) algorithm for streaming anomaly detection, significantly reducing computational overhead.
    \item Validating EILOF's effectiveness through comparative analysis on simulated and real-world datasets, including the Shuttle and Credit Card Fraud datasets.
    \item Investigating the trade-offs in parameter settings (e.g., $k$ and threshold) to optimize performance across various scenarios.
\end{itemize}

The remainder of the paper is organized as follows: Section II provides a brief introduction to both LOF and ILOF. The proposed EILOF algorithm is presented in Section III. In Sections IV and V, we compare the performance of EILOF with standard ILOF using both simulated environments and real-world data. The results showed that EILOF outperforms ILOF as the volume of streaming data increases in various scenarios.

\section{Preliminary}

In this section, we briefly introduce the Local Outlier Factor (LOF) algorithm, and the Incremental LOF (ILOF) algorithm, which is designed for computing LOF scores for data streaming.

\subsection{LOF Algorithm}

In the LOF algorithm, the LOF score of a data point is determined by comparing its density with the densities of its neighbors. The LOF score of a point $p$ is computed based on its Local Reachability Density (LRD), which can be viewed as an approximate kernel density estimate for the point $p$ \cite{schubert2014generalized}.

For a given $k$, the LRD of a point \( p \) is given by:
\begin{equation}
\text{LRD}(p) =  \left(\frac{1}{k} \sum_{o \in N(p, k)} \text{reach-distance}(p, o)\right)^{-1},
\label{eq:lrd}
\end{equation}
where \( N(p, k) \) denotes the set of \( k \)-nearest neighbors of \( p \),
and $\text{reach-distance}(p, o)$ is the $k$-reachability distance between two points \( p \) and \( o \). We will omit $k$ when it does not cause ambiguity. The reachability distance between two points \( p \) and \( q \) is defined as:
\begin{equation}
\text{reach-distance}(p, q) = \max(d(p, q), k\text{-dist}(q)),
\label{reach-dis}
\end{equation}
where \( d(p, q) \) is the Euclidean distance between points \( p \) and \( q \), and \( k\text{-dist}(q) \) is the distance from \( q \) to its \( k \)-th nearest neighbor. The use of \(\max()\) in the reachability distance calculation ensures that the distance measure remains appropriately scaled. In dense areas, where \(d(p, q)\) may be smaller than \(\text{k-dist}(q)\), \(\max()\) prevents the distance from being too small. In sparse areas, where \(d(p, q)\) exceeds \(\text{k-dist}(q)\), \(\max()\) accurately reflects the actual separations. This approach prevents very small distances from inflating the LRD values, thereby ensuring a more reliable calculation. It is important to note that the reachability distances between two points are not symmetric, meaning $\text{reach-distance}(p, q) \neq \text{reach-distance}(q, p)$.

After define the LRD, the LOF score of a point \( p \) is then calculated as:
\begin{equation}
\text{LOF}(p) = \frac{1}{k} \sum_{o \in N(p, k)} \frac{\text{LRD}(o)}{\text{LRD}(p)}\ ,
\label{eq:lof_pc}
\end{equation}
which is the ratio of the average LRD of the $k$-nearest neighbors of $p$ to the LRD of $p$. A point is considered an outlier if its LOF score exceeds a certain threshold, indicating that its local density is much lower than that of its $k$-neighbors. In practice, the threshold is often determined by the proportion of outliers. It is worth emphasizing that the performance of LOF-type algorithms is sensitive to the choice of $k$, which depends on the pattern of the dataset \cite{cao2021scalable}. One might need to conduct several numerical experiments to select the optimal $k$.

\subsection{ILOF Algorithm}
The LOF algorithm requires recalculating $k$-nearest neighbors (k-NN) and density estimates for the entire dataset whenever new data is added. This makes it computationally prohibitive for real-time anomaly detection and limits its adaptability to evolving data distributions, such as concept drift in data streams.

To address the computational inefficiency of the LOF algorithm in handling data streaming, an incremental version of the LOF algorithm (ILOF) was proposed in\cite{pokrajac2007incremental}. Instead of updating the LOF scores of the entire dataset when new data streams in, ILOF only recalculates the reachability distances and LRDs for the points that are directly impacted by the addition. These include the newly added point and any points for which the new data point is among their $k$-nearest neighbors, as well as the $k$-nearest neighbors of those points. For example, assume that point $B$ is a neighbor of point \( A \), and a new point \( E \) is also a neighbor of point \( A \), the reachability distance from point $B$ to point $A$ may also change. This is because the addition of \( E \) can alter the k-th nearest distance of \( A \), which in turn may change the reachability distance from point $B$ to point $A$ due to the definition (\ref{reach-dis}). Once the reachability distances are updated, the next step in the ILOF algorithm is to recalculate the LRD and LOF scores for the affected points (as shown in Algorithm \ref{alg:ilof}). By only adjusting the necessary components of the reachability distance matrix and the affected points of LOF scores, the ILOF algorithm achieves equivalent detection performance as the iterated static LOF algorithm, which re-applies the LOF algorithm from scratch after insertion of new data points, but with significantly reduced computational time. 

\begin{algorithm}
\caption{Incremental LOF Update}
\label{alg:ilof}
\begin{algorithmic}[1]
\Require
\Statex $S \subseteq \mathbb{R}^D$: The current dataset.
\Statex $p_c \in \mathbb{R}^D$: The new data point to be added.
\Statex $k \in \mathbb{Z}^+$: The number of nearest neighbors to consider.
\Ensure
\Statex $\text{LOF}$: Local outlier factors for $p_c$ and affected points in $S$.

\Procedure{Incremental LOF Update}{$S, k, p_c$}
    \State Insert $p_c$ into the dataset $S$
    \State Compute $k$-nearest neighbors of $p_c$, denoted as $N(p_c, k)$, and calculate $k$-distance($p_c$)
    \For{each $p_i \in N(p_c, k)$}
        \State Compute the reachability distance $\text{reach-dist}(p_c, p_i)$
    \EndFor

    \State $S_{\text{update}} \gets $ All points in $S$ where $p_c$ is one of their $k$-nearest neighbors

    \For{each $p_i \in S_{\text{update}}$ and $p_j \in N(p_i, k)$}
        \State Update $k$-distance($p_i$) and $\text{reach-dist}(p_j, p_i)$
        \If{$p_i \in N(p_j, k)$}
            \State $S_{\text{update}} \gets S_{\text{update}} \cup \{p_j\}$
        \EndIf
    \EndFor

    \For{each $p_i \in S_{\text{update}}$}
    \State Recalculate the local reachability density (LRD) for $p_i$
    \State Update the LOF of $p_i$ using the set of points for which $p_i$ is a $k$-nearest neighbor (reverse neighbors)
\EndFor

    \State Compute LRD and LOF of $p_c$ 
    \State \Return $\text{LOF}$
\EndProcedure
\end{algorithmic}
\end{algorithm}

Despite being designed for incremental data processing, ILOF can still incur significant computational costs, particularly when the size of the dataset grows rapidly or has high dimensionality. Additionally, the performance of LOF-type algorithms is highly dependent on the appropriate selection of the parameter $k$, the number of nearest neighbors used to compute the LOF score, and extensive tuning is often required. 
In the case of the ILOF algorithm, if the new data stream exhibits a significantly different pattern from the original data points, even the original points can be heavily impacted, leading to imprecise detection results due to changes in the optimal value of $k$. To address these issues, we propose a new algorithm that reduces computational costs and is less sensitive to parameter selection. Moreover, if an existing data point is definitively identified as an outlier, its LOF score will remain unchanged, even if it becomes part of new data clusters, thereby preserving the properties of the original dataset when new data streams are introduced.

\section{EILOF:  Efficient Incremental Local Outlier Factor}
In this section, we introduce the Efficient Incremental Local Outlier Factor (EILOF) algorithm, which only computes the LOF scores of new points without altering the LOF scores of existing points. The new algorithm gains
significant computational efficiency without a full recalculation of all reachability distances and LOF scores of all affected points. A comprehensive comparison between EILOF and ILOF reachability distance calculations is presented to highlight their differences.

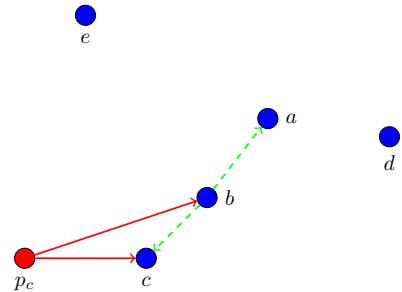
\begin{figure}[htbp]
    \centering
    \resizebox{0.3\textwidth}{!}{
        \begin{tikzpicture}
            \node[draw, circle, fill=red, label=below:$p_c$] (pc) at (0, 0) {};
            \node[draw, circle, fill=blue, label=right:$a$] (a) at (4, 2.3) {};
            \node[draw, circle, fill=blue, label=right:$b$] (b) at (3, 1) {};
            \node[draw, circle, fill=blue, label=below:$c$] (c) at (2, 0) {};
            \node[draw, circle, fill=blue, label=below:$d$] (d) at (6, 2) {};
            \node[draw, circle, fill=blue, label=below:$e$] (e) at (1, 4) {};
            
            \draw[->, red, thick] (pc) -- (b);
            \draw[->, red, thick] (pc) -- (c);
            \draw[->, green, dashed, thick] (b) -- (a);
            \draw[->, green, dashed, thick] (b) -- (c);
        \end{tikzpicture}
    }
    \caption{Graphical representation showing the insertion of a new point $p_c$ and its nearest neighbors. The dashed arrows indicate that $p_c$ is not a nearest neighbor of $b$.}
    \label{fig:graph_representation}
\end{figure}

Specifically, for a new point \( p_c \), the algorithm computes the Euclidean distances to all existing points to identify \( p_c \)'s \( k \)-nearest neighbors. Among these \( k \)-nearest neighbors, we only update their reachability distances to $p_c$ if \( p_c \) is also in the \( k \)-nearest neighbors of these points. This is the key difference between EILOF and ILOF. EILOF ensures that we do not recompute existing reachability distances within the current dataset and avoids updating unnecessary reachability distances from \( p_c \)'s nearest neighbors to it. Furthermore, it does not update the LRD for a data point that is a neighbor of \( p_c \) unless \( p_c \) is one of its \( k \)th nearest neighbors, thereby optimizing the process.
As illustrated, we consider a simple example shown in Figure \ref{fig:graph_representation}, where  $k = 2$  and  $p_c$ is the new point being added to the dataset. The first step is to determine the 2 nearest neighbors of  $p_c$ , which are points  $b$  and  $c$ (as indicated by the solid red arrows in the diagram). The algorithm then computes the reachability distances from  $p_c$  to  $b$  and  $c$ , corresponding to the entries  $p_{c2}$  and  $p_{c3}$ in the reachability distance matrix, as shown in Equation~\ref{eq:reachability_matrix} below.

\begin{equation}
\text{EILOF Reachability Matrix} =
\begin{bmatrix}
    a_{1} & a_{2} & a_{3} & a_{4} & a_{5} & 0 \\[6pt]
    b_{1} & b_{2} & b_{3} & b_{4} & b_{5} & 0 \\[6pt]
    c_{1} & c_{2} & c_{3} & c_{4} & c_{5} & \textcolor{red}{c_{6}} \\[6pt]
    d_{1} & d_{2} & d_{3} & d_{4} & d_{5} & 0 \\[6pt]
    e_{1} & e_{2} & e_{3} & e_{4} & e_{5} & 0 \\[6pt]
    0 & \textcolor{red}{p_{c2}} & \textcolor{red}{p_{c3}} & 0 & 0 & 0 \\
\end{bmatrix},
\label{eq:reachability_matrix}
\end{equation}

Next, the algorithm checks whether \( p_c \) is also a neighbor of \( b \) and \( c \). As shown in Figure \ref{fig:graph_representation}, point \(c \) and point \( p_c \) are mutual neighbors, meaning that both the row and column corresponding to point \(c \) will need to be updated (entry \( c_{6} \) and \( p_{c3} \)). In contrast, although point \( b \) remains a neighbor of \( p_c \), \( p_c \) is no longer one of point \( b \)’s two closest neighbors (as indicated by the dashed green arrows). Consequently, the row corresponding to point  \( b \) does not require an update, and we only need to update the entry corresponding to  \(p_{c2}\). When updating the reachability distances as a new data point is added, the dimension of the reachability distance matrix must be increased by one (Algorithm 2, step 6). We can also see the algorithm updates \( k \) entries in the new row and no more than \( k \) entries in the new column since the reachability distance is not symmetric. Subsequently, the algorithm updates \(\text{LRD}(p_c)\) by using Eq. (\ref{eq:lrd}). We specifically compute \(\text{LOF}(p_c)\) using Eq. (\ref{eq:lof_pc}), while leaving the LOF scores of all other points unchanged.

The pseudocode of EILOF algorithm for general case is shown in Algorithm \ref{alg:eilof}.
\begin{algorithm}
\caption{Efficient Incremental LOF Update}
\label{alg:eilof}
\begin{algorithmic}[1]
\Require 
\Statex $S \subseteq \mathbb{R}^D$: The current set of data points in $D$-dimensional space.
\Statex $k \in \mathbb{Z}^+$: The number of nearest neighbors to consider.
\Statex $p_c \in \mathbb{R}^D$: The new data point to be added to the set $S$.
\Statex $RDM \in \mathbb{R}^{|S| \times |S|}$: The current reachability distance matrix for $S$.
\Ensure 
\Statex $RDM_{\text{updated}}$: The updated reachability distance matrix.
\Statex $\text{LRD}$: Local reachability densities for each point in $S \cup \{p_c\}$.
\Statex $\text{LOF}$: Local outlier factors for each point in $S \cup \{p_c\}$.

\Procedure{Incremental Update}{$S, k, p_c, RDM$}
    \State $S_{\text{updates}} \gets \{\}$ \Comment{Initialize an empty list to keep track of points needing updates}
    \State $S \gets S \cup \{p_c\}$ \Comment{Include the new data point $p_c$ in the set $S$}
    \State Compute distance matrix for $S$
    \State Identify $k$-nearest neighbors of $p_c$
    \State $RDM_{updated} \gets$ Expand $RDM$ to $(|S|+1) \times (|S|+1)$ by adding a new row and column, initially set to 0
    \For{each $p_j$ in the $k$-nearest neighbors of $p_c$}
        \State Compute reachability distance between $p_c$ and $p_j$
        \State Update $RDM_{updated}$ for the new row \Comment{Update the reachability distance matrix for the new row}
        \If{$p_c$ is in the $k$-nearest neighbors of $p_j$}
            \State Compute reachability distance from $p_j$ to $p_c$
            \State Update $RDM_{updated}$ for the new column \Comment{Update the reachability distance matrix for the new column}
            \State $S_{\text{updates}} \gets S_{\text{updates}} \cup \{p_j\}$ \Comment{Add $p_j$ to the list of points needing LRD updates}
        \EndIf
    \EndFor
    \For{each $p_k$ in $S_{\text{updates}}$}
        \State Update LRD for $p_k$
    \EndFor
    \State Compute LOF for $p_c$
    \State \Return Updated $RDM$, LRD, and LOF for $p_c$
\EndProcedure
\end{algorithmic}
\end{algorithm}

Compared to ILOF, which updates the reachability distances and recalculates the LOF scores for all points whose \(k\)-nearest neighbors include the new data point, EILOF simplifies this process by updating only the reachability distances for points directly affected by the new data point. To illustrate the computational differences, we applied ILOF algorithm to the same setup as depicted in Figure~\ref{fig:graph_representation}. Based on the reachability distance calculation from Eq. (\ref{reach-dis}), when the new point \(p_c\) becomes one of \(c\)'s nearest neighbors, \(k\text{-dist}(c)\) changes. This directly affects \(\text{reach-dist}(p, c)\), requiring updates to the reachability distances from all other points to \(c\).

In this example, ILOF updates additional entries such as \(a_3\), \(b_3\), \(c_3\), \(d_3\), and \(e_3\) (shown in Eq. (\ref{eq:reachability_matrix_ILOF}) below).
\begin{equation}
\text{ILOF Reachability Matrix} = 
\begin{bmatrix}
    a_{1} & a_{2} & \textcolor{red}{a_{3}} & a_{4} & a_{5} & 0 \\[6pt]
    b_{1} & b_{2} & \textcolor{red}{b_{3}} & b_{4} & b_{5} & 0 \\[6pt]
    c_{1} & c_{2} & \textcolor{red}{c_{3}} & c_{4} & c_{5} & \textcolor{red}{c_{6}} \\[6pt]
    d_{1} & d_{2} & \textcolor{red}{d_{3}} & d_{4} & d_{5} & 0 \\[6pt]
    e_{1} & e_{2} & \textcolor{red}{e_{3}} & e_{4} & e_{5} & 0 \\[6pt]
    0 & \textcolor{red}{p_{c2}} & \textcolor{red}{p_{c3}} & 0 & 0 & 0 \\
\end{bmatrix},
\label{eq:reachability_matrix_ILOF}
\end{equation}

In this example, we observe that more elements in the reachability distance matrix receive an additional update in ILOF compared to EILOF, specifically the third column (as shown in Eq. (\ref{eq:reachability_matrix_ILOF})). As the dataset grows larger and the number of neighbors increases, ILOF needs to update all points affected by the new data point. This makes the algorithm inefficient for large datasets or high-dimensional data. In contrast, EILOF does not require finding all points whose $k$-nearest neighborhoods include the new data point, nor the $k$-nearest neighborhoods of those points.

Beyond the differences in reachability distance calculations, the EILOF algorithm only computes the LOF score for the new data point, avoiding recalculation of LOF scores for existing points. This design strikes a balance between computational efficiency and accuracy in LOF calculations. However, it is important to emphasize that the accuracy of precise LOF score calculations is distinct from the accuracy of detection results. Given that datasets inherently contain noise, minor deviations in LOF score computations do not necessarily degrade detection performance. In fact, this approach may often yield better results by reducing overfitting and potentially improving the accuracy of outlier detection.

\section{Simulation Studies}

In this section, we present simulation studies on synthetic data to evaluate the performance of the EILOF algorithm compared to the ILOF algorithm. Specifically, we examine the performance of both algorithms by systematically varying the number of neighbors ($k$) and the size of incremental data points ($m$). The performance is measured by the $F_1$ score, a metric that evaluates a model's performance by considering both precision and recall. The $F_1$ score is particularly useful for imbalanced datasets, where one class significantly outnumbers the other, as it balances the trade-off between false positives and false negatives, making it well-suited for tasks like outlier detection. The $F_1$ score is defined as the harmonic mean of precision and recall, balancing their contributions as follows: 
$$F_1 = 2 \cdot \frac{\text{Precision} \cdot \text{Recall}}{\text{Precision} + \text{Recall}}.$$
The simulation results show that for a fixed $k$, although the $F_1$ score decreases for both algorithms as $m$ increases when $m$ is larger than a certain value, the EILOF algorithm shows more robustness with a slower decline in $F_1$ score. Consequently, the EILOF can outperform the ILOF for large $m$ in the simulated data.

\subsection{Setting of the simulated dataset}

The simulated dataset was created based on a specified outlier proportion that determines the number of outliers in the dataset. We began by utilizing data points drawn from a Gaussian distribution centered around the origin as our baseline setting. Outliers were then generated and distinguished from these baseline points by applying scaling and shifts to the Gaussian distribution.

We combined both the normal and the outlier data into a single dataset and then assigned binary labels: 0 for normal points and 1 for outliers. The dataset comprised a total of 2280 data points, partitioned into an initial set of 1000 data points and an additional set of 1280 data points. This split was strategically chosen to manage computational limits and to simulate a realistic scenario where data is accumulated over time, mirroring real-world data inflows. In the initial set, the proportion of outliers was kept low, and more outliers were gradually introduced in the additional set to reflect the online nature of the problem.

The dataset was designed with a 5\% outlier proportion, a common convention in outlier detection research that reflects realistic scenarios where outliers constitute a minority of the data \cite{lazarevic2005feature}. This proportion ensures that the presence of outliers is significant enough to test the robustness of the LOF type algorithms without overwhelming the majority of normal data. To demonstrate the effects of $k$ and $m$, we focus on a 2D dataset to eliminate other factors, such as the curse of dimensionality \cite{zhang2006detecting, aggarwal2001outlier, bellman1959mathematical}, from influencing the outcome of outlier detection. Additionally, we also compare the performance of both algorithms in a 50-dimensional dataset.

\begin{figure}[htbp]
\centering
\includegraphics[width=\linewidth]{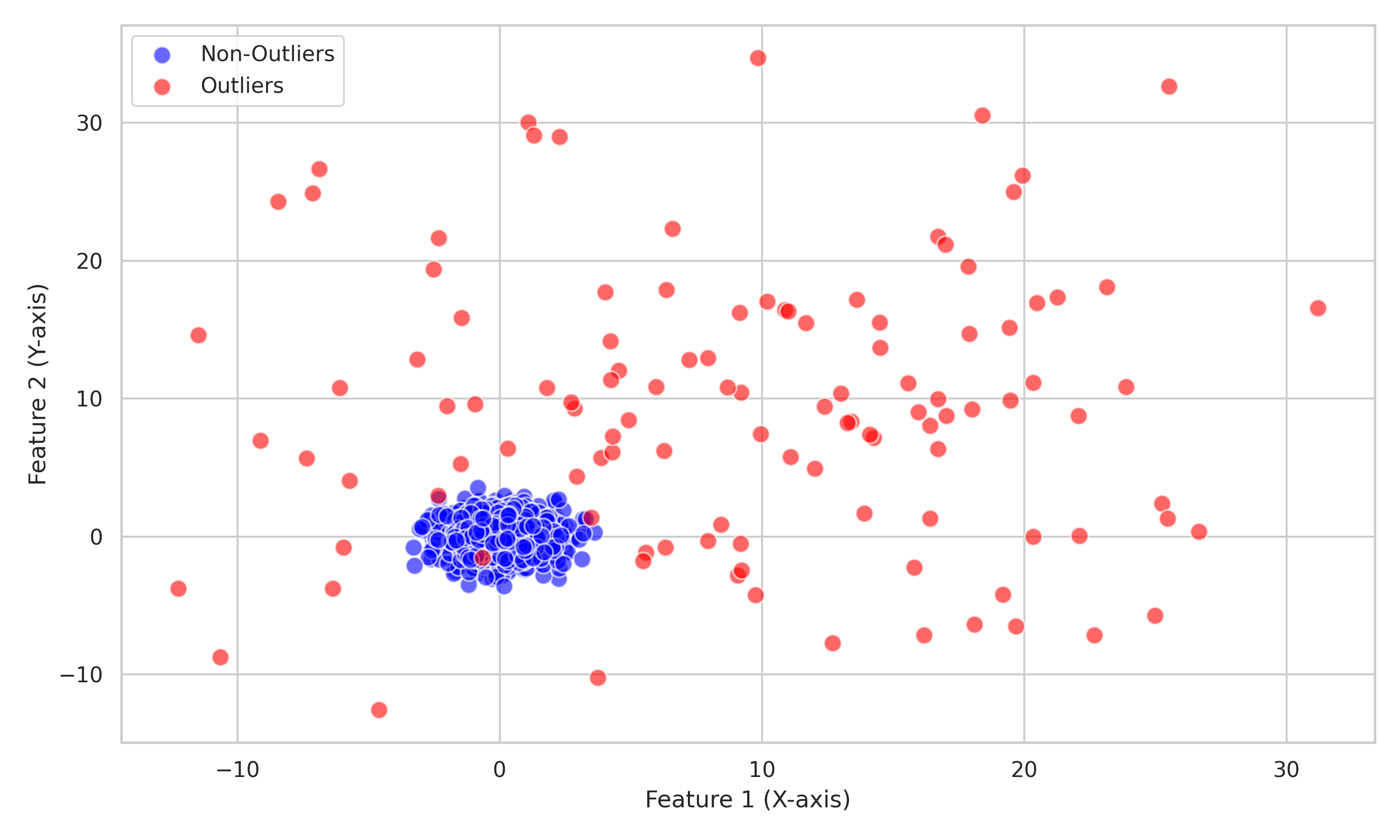}
\caption{Distribution of Simulated Data Points}
\label{fig:distribution}
\end{figure}

The scatter plot in Figure \ref{fig:distribution} illustrates the distribution of the simulated data points in 2D. Data points classified as normal are represented by blue circles, while those identified as outliers are marked with red circles. The plot visually demonstrates the distinction between the two groups: normal data points cluster tightly around the origin, forming a dense core, whereas outliers are more scattered, located farther from the center. This clear visual separation helps illustrate how the LOF algorithm can effectively identify anomalies by assessing local density deviations. The scatter plot provides an intuitive understanding of the dataset’s structure, emphasizing how normals and outliers are distributed for evaluating the LOF algorithm’s performance.

The experimental design focused on simulating a realistic data accumulation process. We selected data point increments in sizes of 1, 5, 10, 20, 40, 80, 160, 320, 640 and 1280 to incrementally challenge the algorithm, assessing its scalability and performance as more data were introduced. These values were chosen to represent a range of typical scenarios in real-world applications, where data streams in at different rates. Smaller increments (e.g., 1, 5, 10, 20, 40) simulate environments with frequent, but smaller, data updates, common in applications like network monitoring and financial fraud detection. Larger increments (e.g., 640, 1280) simulate batch processing scenarios, where data is collected over time and processed periodically, such as in big data analytics and sensor networks. Intermediate values (e.g., 80, 160, 320) provide a balance, allowing us to observe the algorithm’s adaptability and consistency across various data influx rates.

\subsection{\texorpdfstring{Performance of ILOF for different $k$ and $m$}{Performance of ILOF for different k and m}}

\begin{figure}[htbp]
\centering
\includegraphics[width=\linewidth]{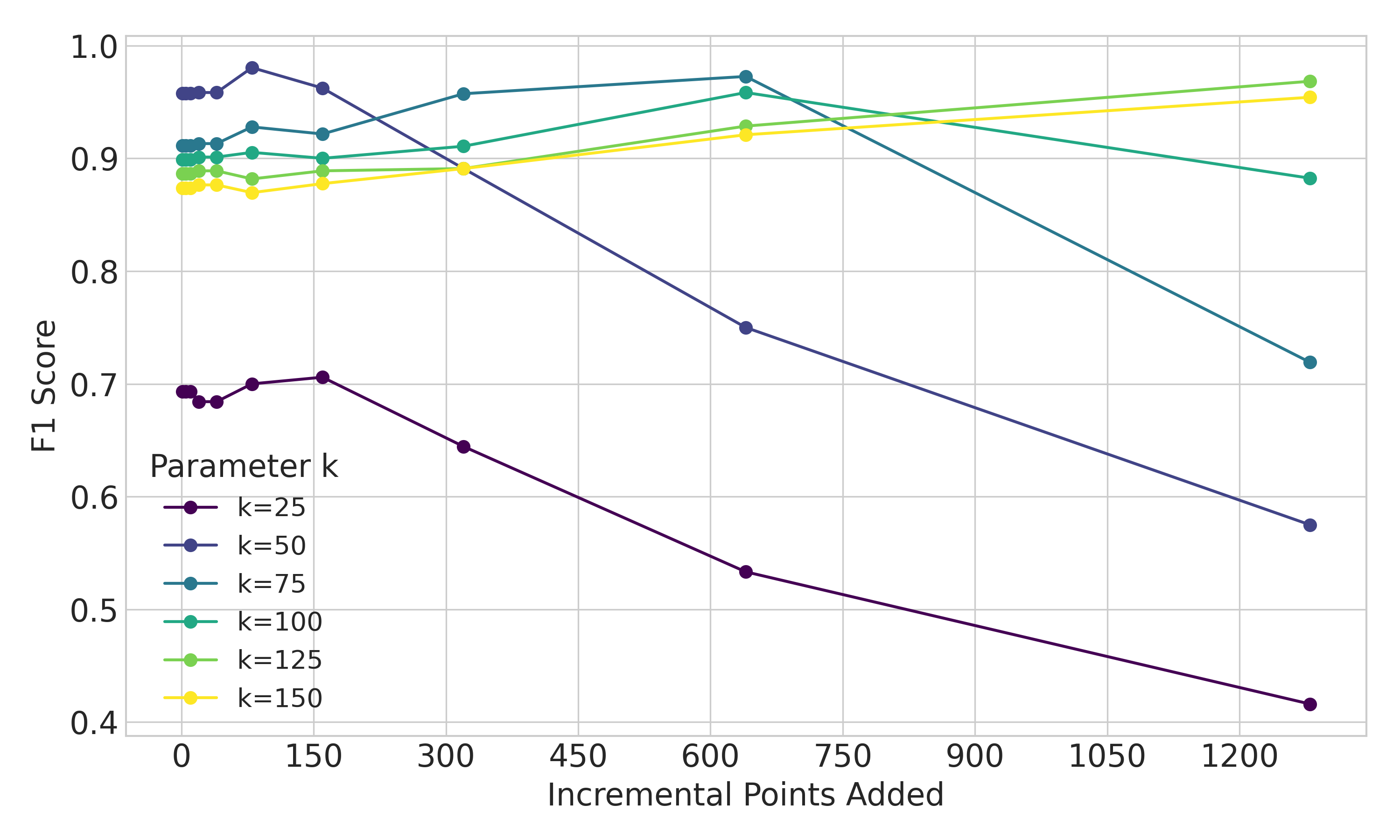}
\caption{$F_{1}$ Score by Test Index for Different $k$ Values in ILOF}
\label{fig:f1score}
\end{figure}

We first examined the performance of the ILOF algorithm, which accurately updates the LOF score for the entire dataset as new data streams in, across various scenarios of incremental data growth.
Figure \ref{fig:f1score} shows the $F_{1}$ score with respect to number of data points added for various fixed $k$ (the number of neighborhoods). 

For smaller values of $k$, the $F_1$ score decreases significantly upon the introduction of new data points. This trend suggests that smaller neighborhood sizes may lack the robustness needed to handle additional data. As $k$ increases, an initial improvement in the $F_1$ score was observed, followed by a subsequent decline. This pattern indicates that an optimal neighborhood size exists, beyond which the algorithm's performance decreases. Moreover, for larger values of $k$, the algorithm demonstrated relatively consistent performance, implying that larger neighborhood sizes provide more stable density estimates and are less affected by the addition of new data points. However, one may expect that the $F_1$ score will decrease when $m$ increases even for large $k$.

The observed peaks in $F_{1}$ score as a function of $k$ and incremental data size 
$m$ in Figure \ref{fig:f1score} can be attributed to the interplay between neighborhood density and outlier sensitivity in the LOF algorithm. Smaller values of $k$ generally lead to tighter neighborhood clusters, making the algorithm more sensitive to local discrepancies in density. Because each neighborhood is small, the algorithm can detect subtle changes and is highly sensitive to how a point's neighbors are distributed. This heightened sensitivity is advantageous up to a certain threshold of data points, beyond which the addition of more data does not translate to better discrimination of outliers, as the neighborhoods become overly dense.

As $k$ increases, the neighborhoods expand to include more points, which initially improves performance by incorporating a broader context and reducing the noise effect from small, local variations. However, beyond a certain point, larger neighborhoods start to dilute the local density differences that are crucial for effective outlier detection, leading to a reduction in $F_1$ scores. The peak performance for each $k$ value represents the optimal balance between enough neighborhood coverage to assess anomalies effectively and sufficient concentration to avoid homogenizing density variations. This peak shifts with changes in $k$, as larger $k$ values require more data to reach an optimal state of neighborhood density.

Furthermore, the diminishing returns from increasing $k$ beyond these peaks suggest that the algorithm reaches a saturation point where additional contextual information no longer contributes to distinguishing outliers effectively, highlighting the trade-off between breadth and precision in anomaly detection settings.

\begin{figure}[htbp]
\centering
\includegraphics[width=\linewidth]{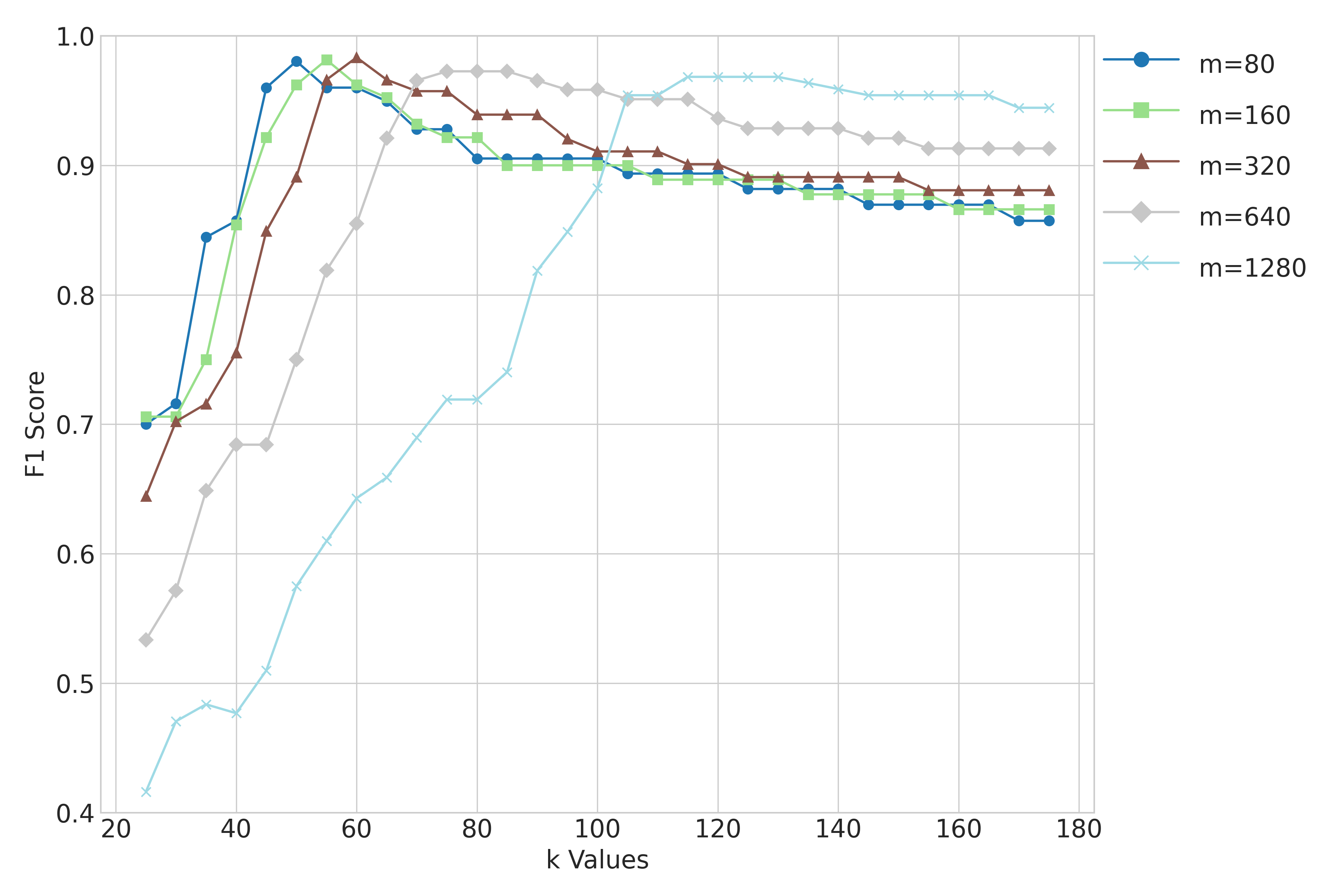}
\caption{$F_{1}$ Score by $k$ for Different Sizes of Incremental Data Points ($m$) in ILOF}
\label{fig:f1_k}
\end{figure}

The observation is further supported by Figure \ref{fig:f1_k}, which shows the $F_1$ score with respect to $k$ for different value of $m$. The corresponding $F_1$ scores for different values of $k$ and $m$ are provided in Table \ref{tab:f1_scores}. The result reveals a distinct pattern: the optimal $F_{1}$ score is attained at different \( k \) values depending on the size of the incrementally added data. The $F_{1}$ score exhibits a non-monotonic relationship with \( k \) for a fixed \( m \), initially increasing before subsequently decreasing as \( k \) grows. Notably, the optimal \( k \) that maximizes $F_{1}$ score demonstrates a positive correlation with \( m \).

For smaller \( k \) values, optimal performance is observed at medium incremental sizes, with effectiveness diminishing at both smaller and larger increments. This suggests that the LOF algorithm's sensitivity to the number of incremental data points is heightened at smaller \( k \) values, with a moderate increment size yielding the most accurate outlier detection.

As \( k \) increases, the trend reverses, with larger incremental sizes leading to improved performance. This indicates that larger \( k \) values benefit from a greater volume of data, enhancing the algorithm's accuracy. The interaction between neighborhood size and data volume highlights the importance of selecting an appropriate \( k \) value, as it significantly impacts the model's precision.

These findings underscore the delicate balance between neighborhood size and data volume in outlier detection algorithms, emphasizing the need for careful calibration of the parameter \( k \) to optimize performance.

Therefore, the parameter \( k \) significantly influences the performance of the Incremental Local Outlier Factor (ILOF) algorithm \cite{li2023outlier, wang2021local}. Identifying and updating the optimal value of \( k \) in data streaming scenarios is an arduous and time-consuming task, primarily due to the high computational complexity inherent in machine learning processes. The optimal \( k \) can vary depending on the number of data points added, underscoring the importance of finding alternative solutions to mitigate the impact of updating the optimal  \(k\)  value in data streaming scenarios.

\subsection{\texorpdfstring{Performance of EILOF for different $k$ and $m$}{Performance of EILOF for different k and m}}
\label{sec:ilof_performance}

\begin{figure}[htbp]
\centering
\includegraphics[width=\linewidth]{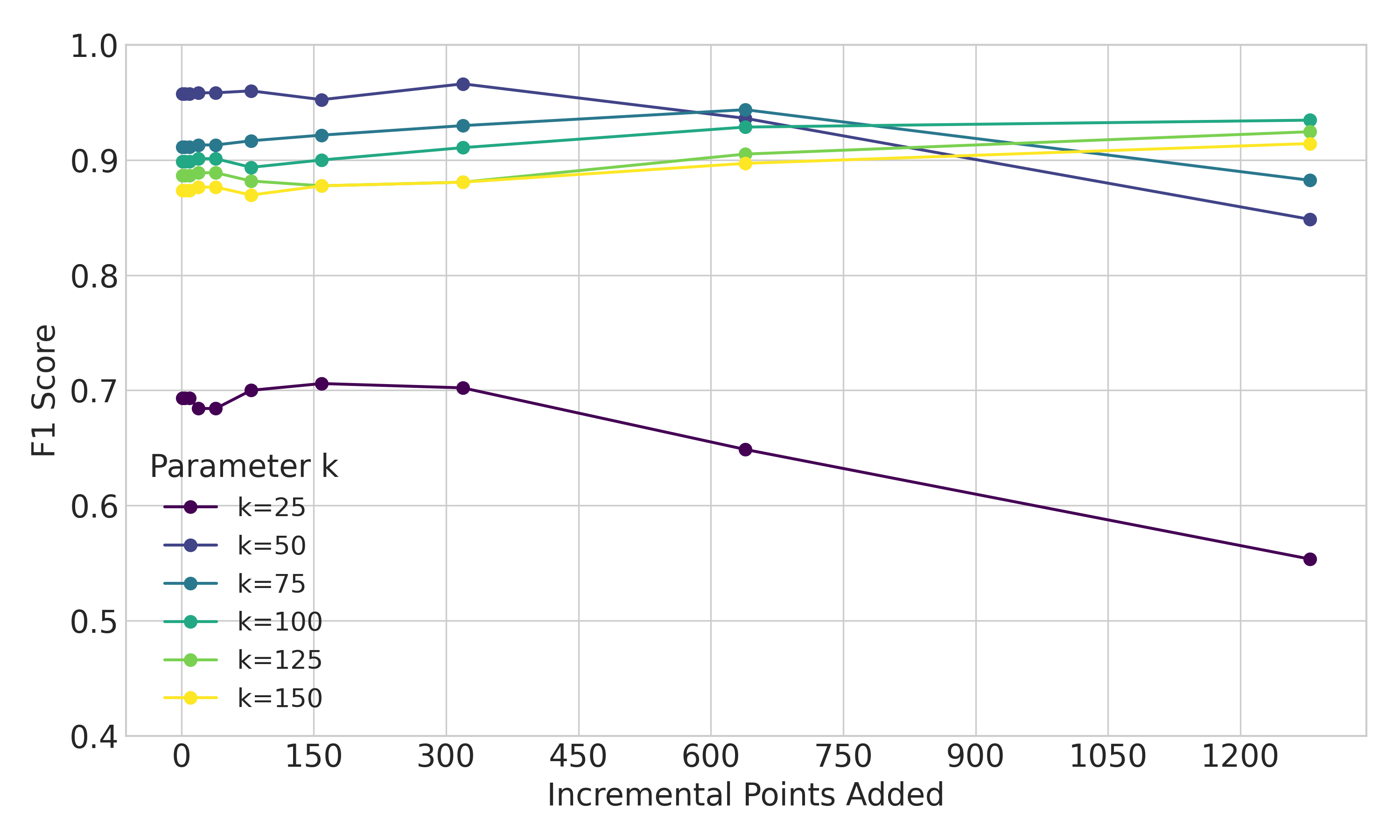}
\caption{$F_{1}$ Score by Test Index for Different $k$ Values in EILOF}
\label{fig:f1score_EILOF}
\end{figure}

\begin{figure}[htbp]
\centering
\includegraphics[width=\linewidth]{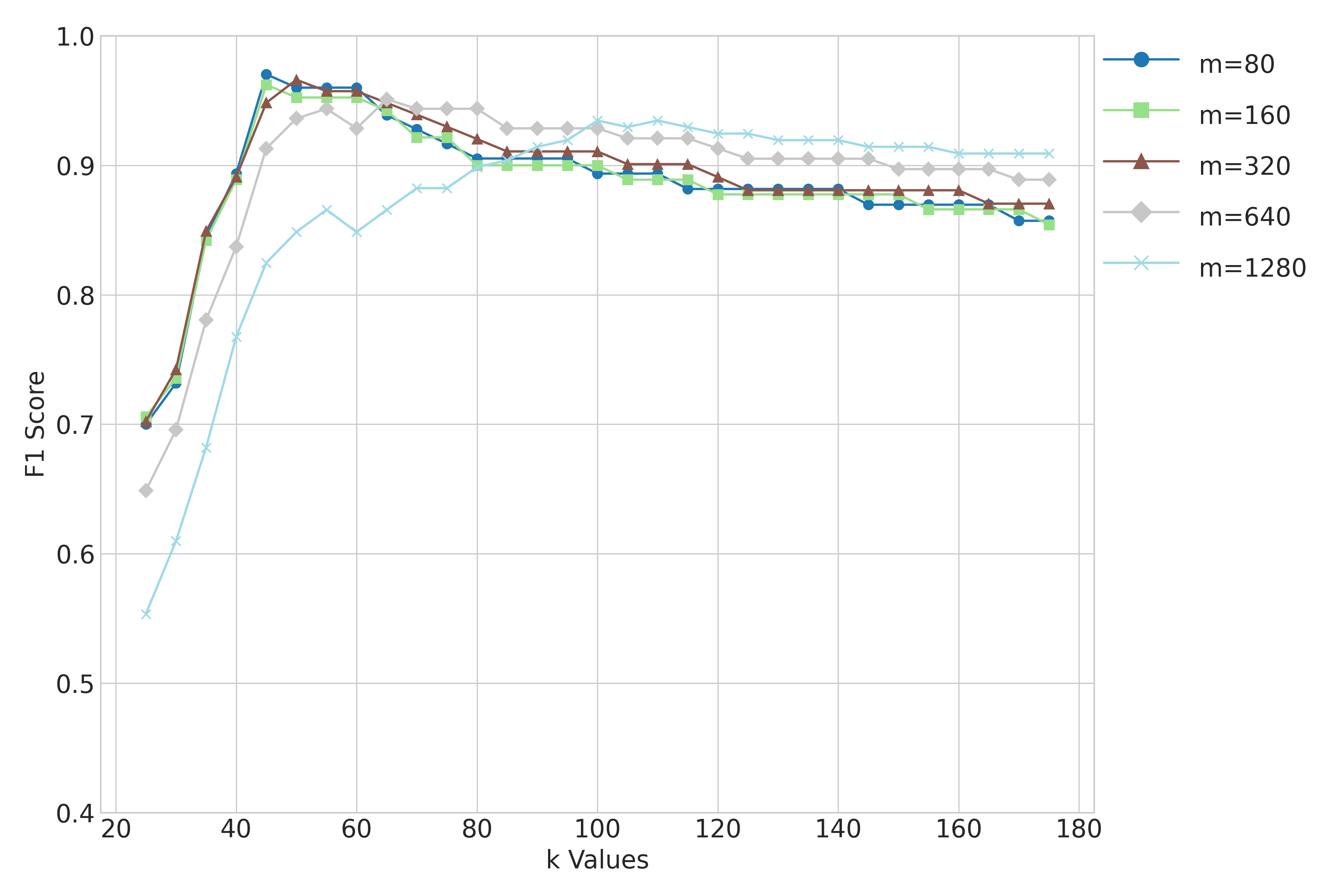}
\caption{$F_{1}$ Score by $k$ for Different Sizes of Incremental Data Points ($m$) in EILOF}
\label{fig:f1_k_EILOF}
\end{figure}

Next, we conducted the same experiment for EILOF with different $k$ and $m$ values to see whether these patterns persist or if any improvements are observed during incremental analysis. Figure \ref{fig:f1score_EILOF} and  Figure \ref{fig:f1_k_EILOF} show the $F_1$ score with respect to $m$ for different values of $k$, and  with respect to $k$ for different values of $m$, in EILOF, respectively.

It can be noticed that the general trend of these curves are similar to those for ILOF.
However, there are subtle difference in compared with ILOF, indicating that EILOF generates more stable results across different scenarios.
For $k = 25$ and $k =50$, although $F_1$ score decreases when $m$ increases, it decreases much slower than ILOF. For larger $k$, the EILOF also shows relatively consistent performance.
We also observe that the optimal $F_1$ score is obtained at different $k$ values for different $m$. This observation is crucial and suggest that EILOF is more robust with respect to the choice of $k$ and changes in $m$.
To further support our conclusion, 
we provide a comprehensive comparison using both simulated and real data, with detailed explanations to follow in subsequent sections.

\subsection{Performance Comparison in Simulation Data}

Building on the insights from the previous section, we now turn to a direct comparison between ILOF and EILOF to further understand the advantages of the proposed algorithm.

\begin{figure}[!h]
\centering
\includegraphics[width=\linewidth]{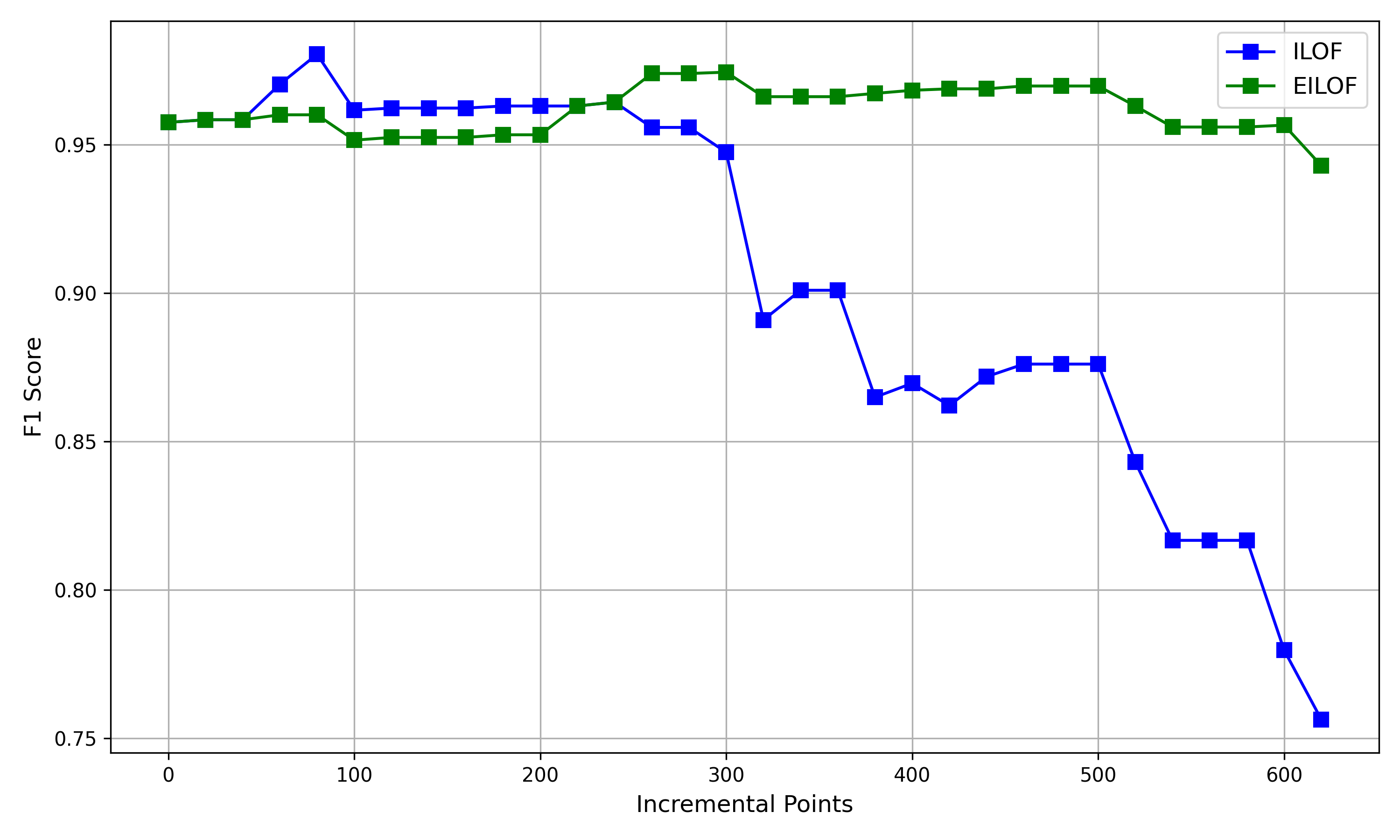}
\caption{Comparison of Incremental $F_{1}$ Scores in $2D$: ILOF vs. EILOF when \( k = 50 \)}
\label{fig:f1_comparison}
\end{figure}

\begin{figure}[!h]
\centering
\includegraphics[width=\linewidth]{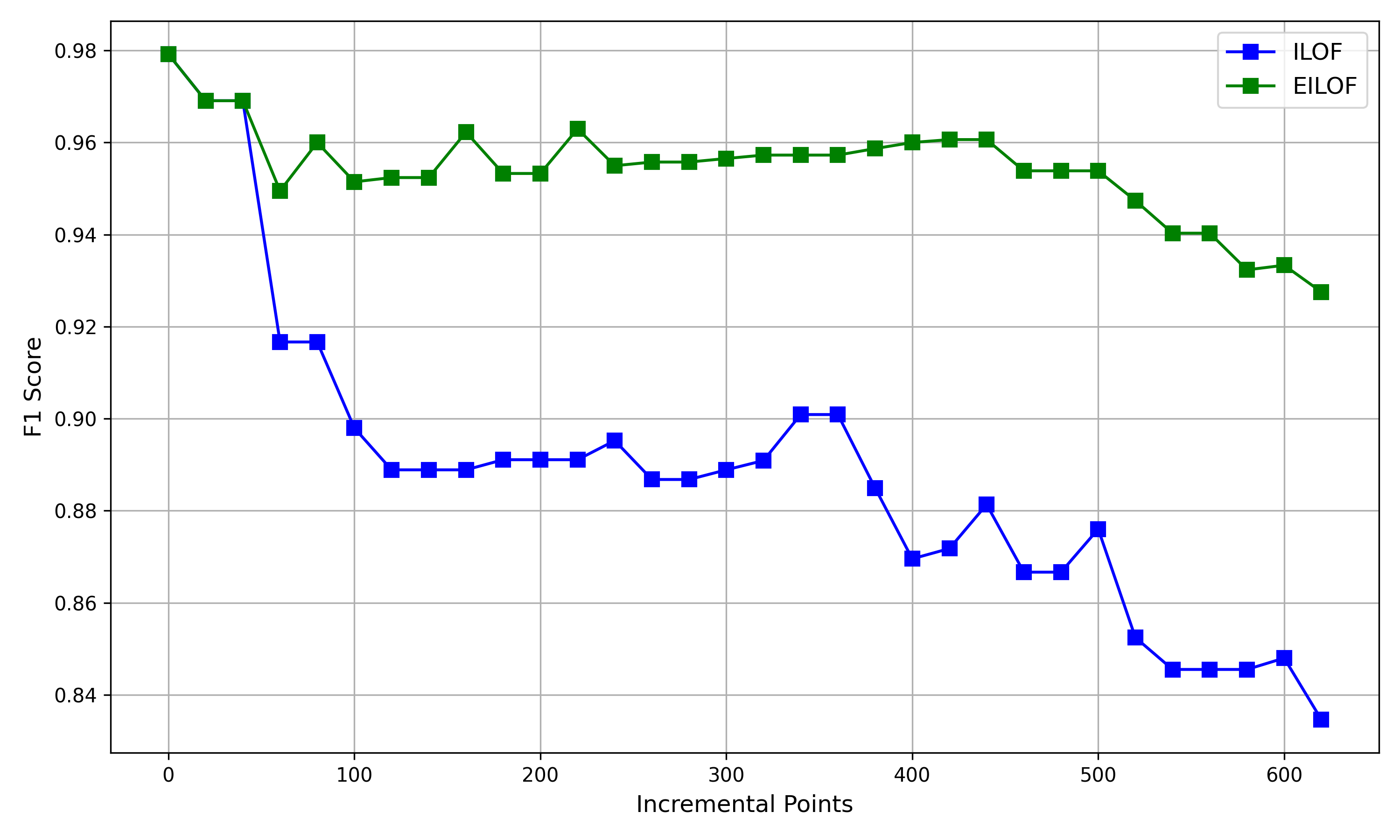}
\caption{Comparison of Incremental $F_{1}$ Scores in $50D$: ILOF vs. EILOF when \( k = 50 \)}
\label{fig:50dimensionf1_comparison}
\end{figure}

Figure \ref{fig:f1_comparison} compares the incremental \( F_1 \) scores of the traditional ILOF algorithm and the proposed EILOF algorithm. Here, we chose $k=50$ and $D=2$ as an example. The graph illustrates the stability and accuracy of the EILOF algorithm in maintaining high \( F_1 \) scores as incremental points are added, compared to the fluctuating performance of the ILOF method. While there may be specific instances where ILOF might perform slightly better, EILOF generally demonstrates strong performance and robustness in dynamic data stream scenarios. In many cases, EILOF’s advantages become more pronounced, highlighting its significant potential for enhancing outlier detection in evolving datasets. Figure \ref{fig:50dimensionf1_comparison} presents a comparative analysis of incremental $F_1$ scores for the case where $k=50$ and $D=50$. While similar behavioral patterns are observed, it is noteworthy that the performance degradation of ILOF occurs earlier than in the 2-dimensional scenario. This phenomenon can be attributed to the increased sparsity of data in higher-dimensional spaces, which consequently leads to reduced robustness of the algorithm.
The earlier onset of performance deterioration in the higher-dimensional case underscores the challenges associated with maintaining algorithmic efficacy as data complexity increases. This observation highlights the importance of considering dimensionality when evaluating and implementing outlier detection algorithms, particularly in scenarios involving high-dimensional datasets.

\begin{table*}[ht]
\centering
\scriptsize 
\begin{adjustbox}{max width=\textwidth}
\begin{tabular}{c|cc|cc|cc|cc|cc|cc|cc|cc|cc}
\toprule
\textbf{k (Neighbors)} & \multicolumn{2}{c|}{\textbf{m = 5}} & \multicolumn{2}{c|}{\textbf{m = 10}} & \multicolumn{2}{c|}{\textbf{m = 20}} & \multicolumn{2}{c|}{\textbf{m = 40}} & \multicolumn{2}{c|}{\textbf{m = 80}} & \multicolumn{2}{c|}{\textbf{m = 160}} & \multicolumn{2}{c|}{\textbf{m = 320}} & \multicolumn{2}{c|}{\textbf{m = 640}} & \multicolumn{2}{c}{\textbf{m = 1280}} \\
 & \textbf{2D} & \textbf{50D} & \textbf{2D} & \textbf{50D} & \textbf{2D} & \textbf{50D} & \textbf{2D} & \textbf{50D} & \textbf{2D} & \textbf{50D} & \textbf{2D} & \textbf{50D} & \textbf{2D} & \textbf{50D} & \textbf{2D} & \textbf{50D} & \textbf{2D} & \textbf{50D} \\
\midrule
25  & 0.6933 & 0.7901 & 0.6933 & 0.7901 & 0.6842 & 0.7654 & 0.6842 & 0.7654 & 0.7000 & 0.7470 & 0.7059 & 0.7356 & 0.6444 & 0.7292 & 0.5333 & 0.6957 & 0.4161 & 0.5926 \\
50  & 0.9574 & 0.9792 & 0.9574 & 0.9792 & 0.9583 & 0.9691 & 0.9583 & 0.9691 & 0.9804 & 0.9167 & 0.9623 & 0.8889 & 0.8909 & 0.8909 & 0.7500 & 0.8372 & 0.5750 & 0.7473\\

75  & 0.9111 & 1.000 & 0.9111 & 1.000 & 0.9130 & 1.000 & 0.9130 & 1.000 & 0.9278 & 1.000 & 0.9216 & 1.000 & 0.9573 & 1.000 & 0.9726 & 0.9865 & 0.7191 & 0.8308\\

100 & 0.8989 & 1.000 & 0.8989 & 1.000 & 0.9011 & 1.000 & 0.9011 & 1.000 & 0.9053 & 1.000 & 0.9000 & 1.000 & 0.9107 & 1.000 & 0.9583 & 1.000 & 0.8824 & 0.9194\\

125 & 0.8864 & 1.000 & 0.8864 & 1.000 & 0.8889 & 1.000 & 0.8889 & 1.000 & 0.8817 & 1.000 & 0.8889 & 1.000 & 0.8909 & 1.000 & 0.9286 & 1.000 & 0.9683 & 0.9867\\

150 & 0.8736 & 1.000 & 0.8736 & 1.000 & 0.8764 & 1.000 & 0.8764 & 1.000 & 0.8696 & 1.000 & 0.8776 & 1.000 & 0.8909 & 1.000 & 0.9209 & 1.000 & 0.9541 & 1.000\\
\bottomrule
\end{tabular}
\end{adjustbox}
\caption{$F_{1}$ Scores for Different Values of \( k \) (Number of Neighbors) and \( m \) (Points Added) across 2 and 50 dimensions for ILOF.}
\label{tab:f1_scores}
\end{table*}

\begin{table*}[ht]
\centering
\scriptsize 
\begin{adjustbox}{max width=\textwidth}
\begin{tabular}{c|cc|cc|cc|cc|cc|cc|cc|cc|cc}
\toprule
\textbf{k (Neighbors)} & \multicolumn{2}{c|}{\textbf{m = 5}} & \multicolumn{2}{c|}{\textbf{m = 10}} & \multicolumn{2}{c|}{\textbf{m = 20}} & \multicolumn{2}{c|}{\textbf{m = 40}} & \multicolumn{2}{c|}{\textbf{m = 80}} & \multicolumn{2}{c|}{\textbf{m = 160}} & \multicolumn{2}{c|}{\textbf{m = 320}} & \multicolumn{2}{c|}{\textbf{m = 640}} & \multicolumn{2}{c}{\textbf{m = 1280}} \\
 & \textbf{2D} & \textbf{50D} & \textbf{2D} & \textbf{50D} & \textbf{2D} & \textbf{50D} & \textbf{2D} & \textbf{50D} & \textbf{2D} & \textbf{50D} & \textbf{2D} & \textbf{50D} & \textbf{2D} & \textbf{50D} & \textbf{2D} & \textbf{50D} & \textbf{2D} & \textbf{50D} \\
\midrule
25  & 0.6933 & 0.7901 & 0.6933 & 0.7901 & 0.6842 & 0.7805 & 0.6842 & 0.7805 & 0.7000 & 0.7765 & 0.7059 & 0.7640 & 0.7021 & 0.7800 & 0.6486 & 0.7395 & 0.5535 & 0.6328 \\

50  & 0.9574 & 0.9792 & 0.9574 & 0.9792 & 0.9583 & 0.9691 & 0.9583 & 0.9691 & 0.9600 & 0.9600 & 0.9524 & 0.9623 & 0.9661 & 0.9573 & 0.9362 & 0.9286 & 0.8485 & 0.7895 \\

75  & 0.9111 & 1.000 & 0.9111 & 1.000 & 0.9130 & 1.000 & 0.9130 & 1.000 & 0.9167 & 1.000 & 0.9216 & 1.000 & 0.9298 & 1.000 & 0.9437 & 1.000 & 0.8824 & 0.9100 \\

100 & 0.8989 & 1.000 & 0.8989 & 1.000 & 0.9011 & 1.000 & 0.9011 & 1.000 & 0.8936 & 1.000 & 0.9000 & 1.000 & 0.9107 & 1.000 & 0.9286 & 1.000 & 0.9346 & 0.9493 \\

125 & 0.8864 & 1.000 & 0.8864 & 1.000 & 0.8889 & 1.000 & 0.8889 & 1.000 & 0.8817 & 1.000 & 0.8776 & 1.000 & 0.8807 & 1.000 & 0.9051 & 1.000 & 0.9245 & 0.9821 \\

150 & 0.8736 & 1.000 & 0.8736 & 1.000 & 0.8764 & 1.000 & 0.8764 & 1.000 & 0.8696 & 1.000 & 0.8776 & 1.000 & 0.8807 & 1.000 & 0.8971 & 1.000 & 0.9143 & 0.9589 \\

\bottomrule
\end{tabular}
\end{adjustbox}
\caption{$F_{1}$ Scores for Different Values of \( k \) (Number of Neighbors) and \( m \) (Points Added) across 2 and 50 dimensions for EILOF.}
\label{tab:f1_scores_new}
\end{table*}

To provide a more comprehensive view of EILOF’s performance, we compared its performance to ILOF across different values of $k$ and varying numbers of added points, as shown in Table \ref{tab:f1_scores} and \ref{tab:f1_scores_new}. Notice that the EILOF algorithm can always perform better in the case of relatively larger $m$ compared to $k$. In fact, for a fixed $k$, as $m$ increases, the performance of ILOF worsens because the fixed $k$ tends to be relatively smaller as the sample size increases, making it a less suitable choice, while in the mean time, EILOF maintains good performance. For larger values of $k$, this decreasing trend appears only after a greater increase in $m$, but the overall pattern is similar. So we could observe that ILOF may perform better than EILOF in some senerios where $k$ is relatively large compare to $m$ but the difference is not obvious. In contrast, the advantages of EILOF in the scenerios where $m$ is relatively large compared to $k$ are obvious. Therefore, we conclude that EILOF is a more robust and efficient algorithm in terms of both computational costs and detection performance.

\section{Performance Comparison in Real Data}

To further validate the robustness and applicability of the proposed EILOF algorithm, we compared its performance against ILOF algorithm on two real-world datasets: the Shuttle dataset from the UCI Machine Learning Repository \cite{misc_statlog_(shuttle)_148} and the Credit Card Fraud dataset from Kaggle \cite{dal_pozzolo_credit_card_fraud}. These datasets were selected to evaluate EILOF under diverse scenarios. The Shuttle dataset, with its structured features and randomized order of examples, is a controlled benchmark for testing anomaly detection algorithms in static scenarios. In contrast, the Credit Card Fraud dataset, characterized by its sequential nature, simulates real-world transaction-based data, making it suitable for evaluating streaming and time-dependent anomaly detection. This comparison demonstrates EILOF's adaptability and effectiveness across different types of data and real-world applications.

\subsection{Shuttle Dataset}
The original Shuttle dataset contains 20\% outliers (labels 2, 3, 4, 5, 6, and 7). To simplify the outlier detection process, we removed the largest outlier group (label 4), a common practice in similar studies \cite{abe2006outlier, 4781136, DING201312}. The remaining outlier labels (2, 3, 5, and 7) were combined into a single outlier class, reducing the outlier proportion to 7\% \cite{tan2011fast}. The dataset contains 7 features and 49,097 observations.

To adapt the dataset to a data streaming problem, we selected the first 1,640 data points, using the first 1,000 observations as static data and the remaining 640 as streaming data. We then evaluated the performance of the Efficient Incremental Local Outlier Factor (EILOF) and Incremental Local Outlier Factor (ILOF) methods under various thresholds (5\%, 7\%, and 10\%) and varying $k$ values (50, 100, 150).

To adapt the dataset for a data streaming problem, we selected the first 1,640 data points, using the first 1,000 observations as static data and the remaining 640 as streaming data. We then compared the performance of the Efficient Incremental Local Outlier Factor (EILOF) and Incremental Local Outlier Factor (ILOF) methods.

\begin{figure}[!h]
    \centering
    \begin{subfigure}[b]{0.4\textwidth}
        \centering
        \includegraphics[width=\linewidth]{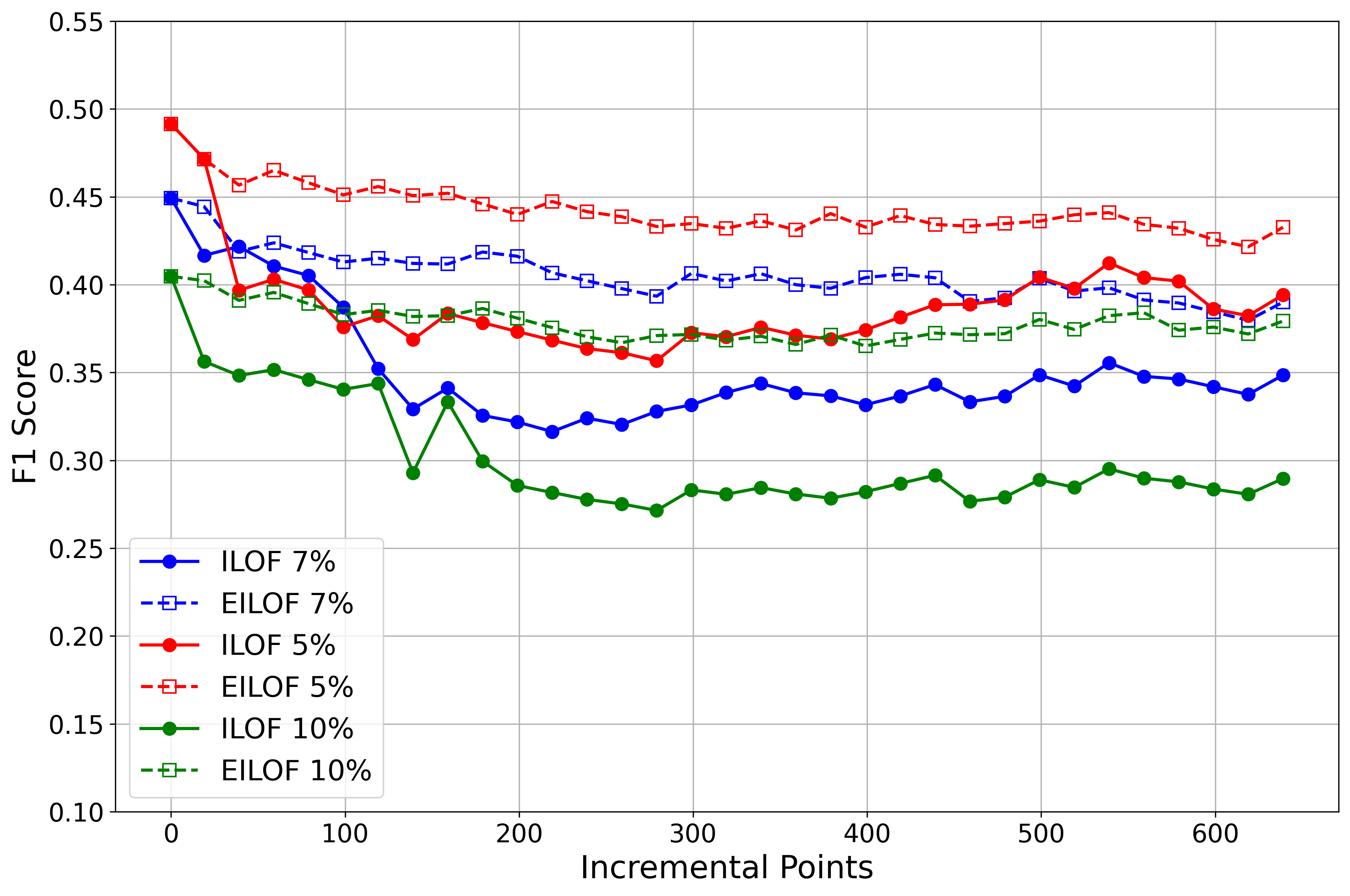}
        \caption{$k = 50$}
        \label{fig:graph1}
    \end{subfigure}
    \hfill
    \begin{subfigure}[b]{0.4\textwidth}
        \centering
        \includegraphics[width=\linewidth]{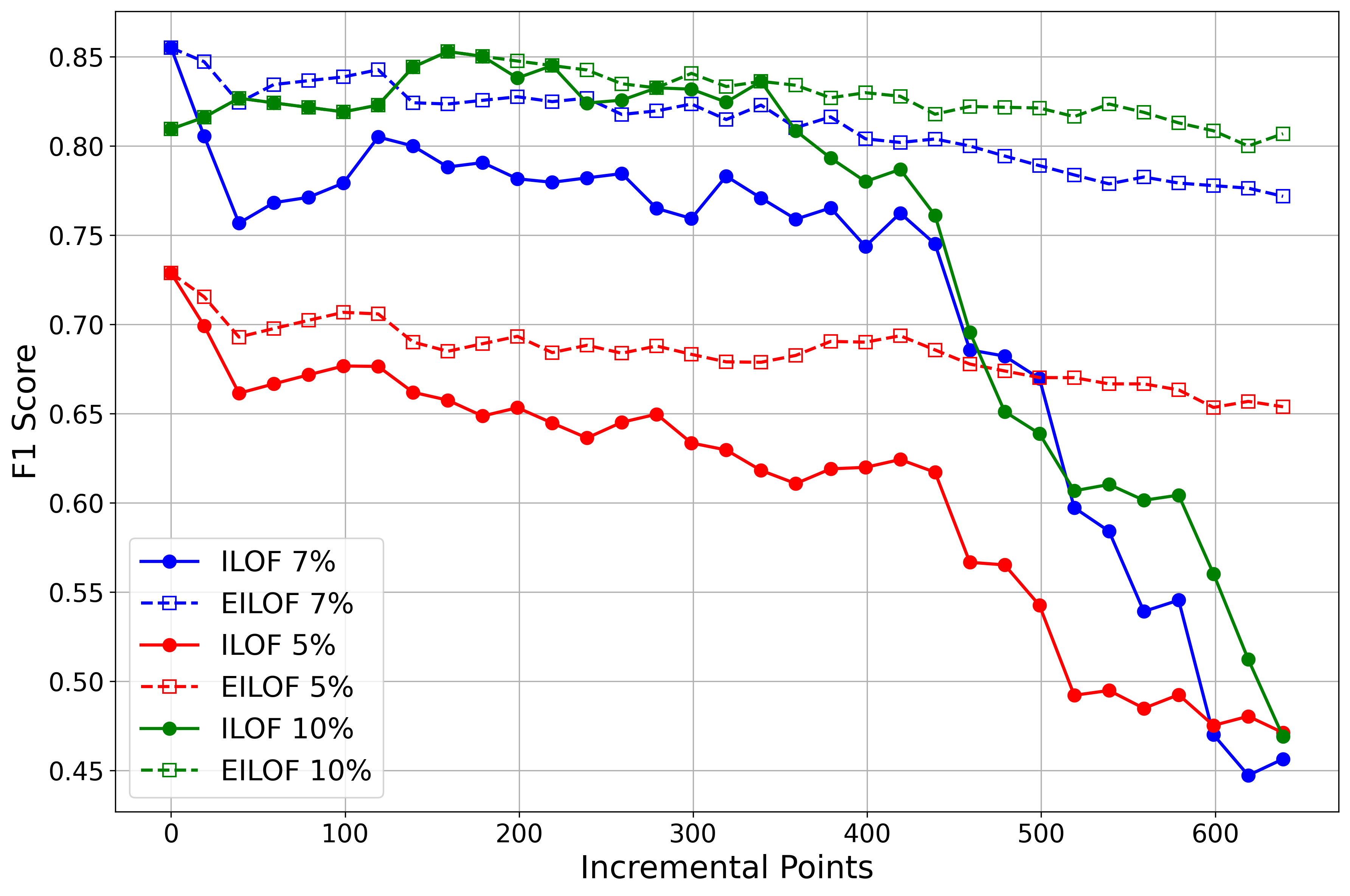} 
        \caption{$k = 100$}
        \label{fig:graph2}
    \end{subfigure}
    \hfill
    \begin{subfigure}[b]{0.4\textwidth}
        \centering
        \includegraphics[width=\linewidth]{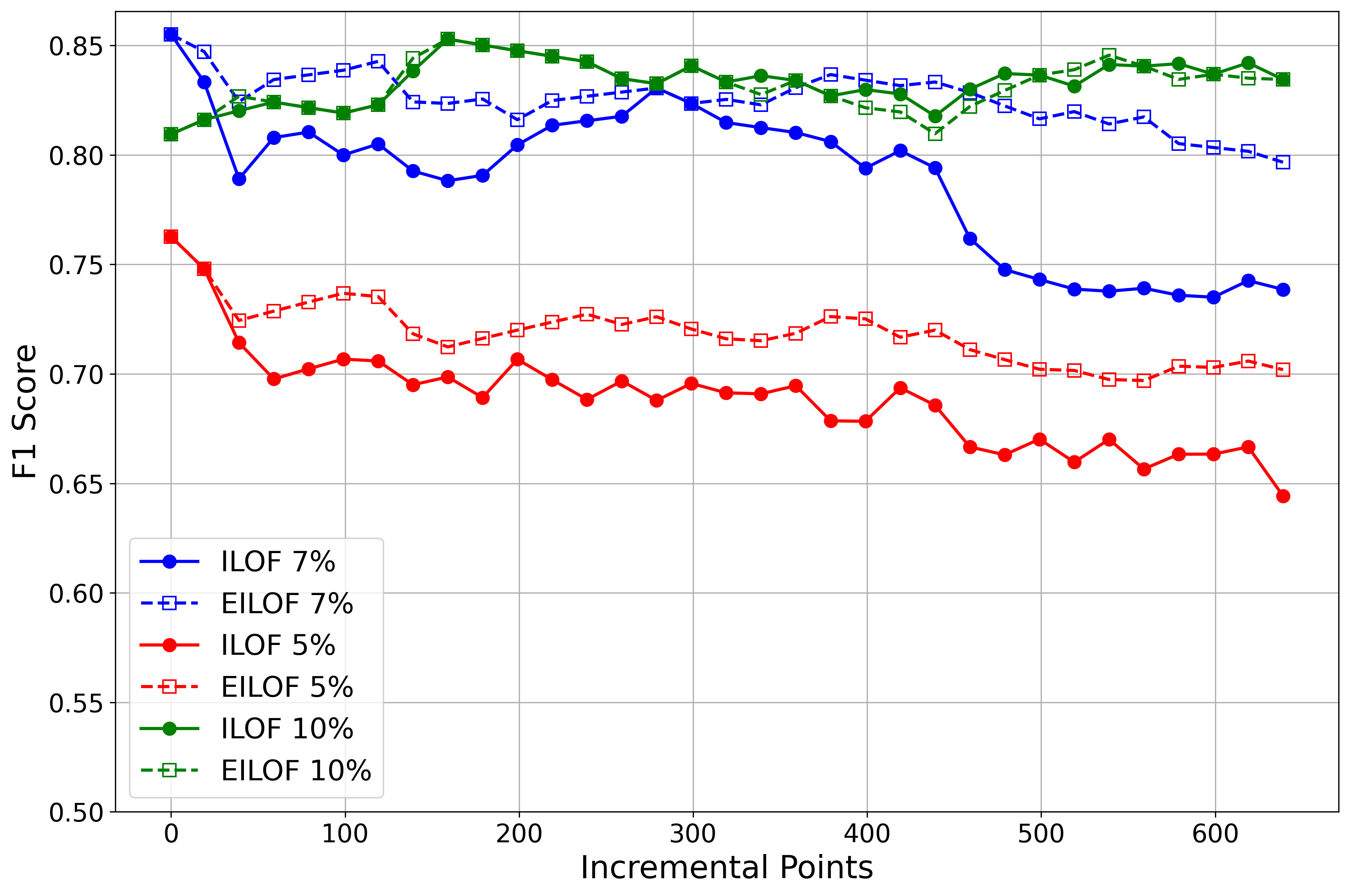} 
        \caption{$k = 150$}
        \label{fig:graph3}
    \end{subfigure}
    \caption{Comparison of EILOF and ILOF performance on Shuttle dataset. The $F_1$ scores are plotted against incremental points. Key points are selected every 20th increment to visualize the performance trend. (a) $k = 50$, (b) $k = 100$, (c) $k = 150$.}
    \label{fig:three_graphs}
\end{figure}

\begin{table*}[ht]
\centering
\Large 
\begin{adjustbox}{max width=\textwidth}
\begin{tabular}{c|ccc|ccc|ccc|ccc|ccc|ccc|ccc|ccc}
\toprule
\textbf{k (Neighbors)} & \multicolumn{6}{c|}{\textbf{m = 5}} & \multicolumn{6}{c|}{\textbf{m = 10}} & \multicolumn{6}{c|}{\textbf{m = 20}} & \multicolumn{6}{c}{\textbf{m = 40}} \\
 & \multicolumn{3}{c|}{\textbf{ILOF}} & \multicolumn{3}{c|}{\textbf{EILOF}} & \multicolumn{3}{c|}{\textbf{ILOF}} & \multicolumn{3}{c|}{\textbf{EILOF}} & \multicolumn{3}{c|}{\textbf{ILOF}} & \multicolumn{3}{c|}{\textbf{EILOF}} & \multicolumn{3}{c|}{\textbf{ILOF}} & \multicolumn{3}{c}{\textbf{EILOF}} \\
 & \textbf{5\%} & \textbf{7\%} & \textbf{10\%} & \textbf{5\%} & \textbf{7\%} & \textbf{10\%} & \textbf{5\%} & \textbf{7\%} & \textbf{10\%} & \textbf{5\%} & \textbf{7\%} & \textbf{10\%} & \textbf{5\%} & \textbf{7\%} & \textbf{10\%} & \textbf{5\%} & \textbf{7\%} & \textbf{10\%} & \textbf{5\%} & \textbf{7\%} & \textbf{10\%} & \textbf{5\%} & \textbf{7\%} & \textbf{10\%} \\
\midrule
50  & {0.4874} & {0.4460} & {0.3905} & {0.4874} & {0.4460} & {0.4024} & {0.4833} & {0.4143} & {0.3765} & {0.4833} & {0.4429} & {0.4000} & {0.4878} & {0.4166} & {0.3563} & {0.4715} & {0.4444} & {0.4023} & {0.3937} & {0.4054} & {0.3464} & {0.4567} & {0.4189} & {0.3911} \\
100 & {0.7227} & {0.8633} & {0.8047} & {0.7227} & {0.8633} & {0.8047} & {0.7167} & {0.8429} & {0.8000} & {0.7167} & {0.8571} & {0.8000} & {0.6992} & {0.8056} & {0.8161} & {0.7154} & {0.8472} & {0.8161} & {0.6614} & {0.7568} & {0.8268} & {0.6929} & {0.8243} & {0.8268} \\
150 & {0.7731} & {0.8633} & {0.8047} & {0.7731} & {0.8633} & {0.8047} & {0.7667} & {0.8571} & {0.8000} & {0.7667} & {0.8571} & {0.8000} & {0.7480} & {0.8333} & {0.8161} & {0.7480} & {0.8472} & {0.8161} & {0.6929} & {0.7838} & {0.8268} & {0.7244} & {0.8243} & {0.8268} \\
\midrule
\multicolumn{1}{c}{} & \multicolumn{6}{c|}{\textbf{m = 80}} & \multicolumn{6}{c|}{\textbf{m = 160}} & \multicolumn{6}{c|}{\textbf{m = 320}} & \multicolumn{6}{c}{\textbf{m = 640}} \\
 & \multicolumn{3}{c|}{\textbf{ILOF}} & \multicolumn{3}{c|}{\textbf{EILOF}} & \multicolumn{3}{c|}{\textbf{ILOF}} & \multicolumn{3}{c|}{\textbf{EILOF}} & \multicolumn{3}{c|}{\textbf{ILOF}} & \multicolumn{3}{c|}{\textbf{EILOF}} & \multicolumn{3}{c|}{\textbf{ILOF}} & \multicolumn{3}{c}{\textbf{EILOF}} \\
 & \textbf{5\%} & \textbf{7\%} & \textbf{10\%} & \textbf{5\%} & \textbf{7\%} & \textbf{10\%} & \textbf{5\%} & \textbf{7\%} & \textbf{10\%} & \textbf{5\%} & \textbf{7\%} & \textbf{10\%} & \textbf{5\%} & \textbf{7\%} & \textbf{10\%} & \textbf{5\%} & \textbf{7\%} & \textbf{10\%} & \textbf{5\%} & \textbf{7\%} & \textbf{10\%} & \textbf{5\%} & \textbf{7\%} & \textbf{10\%} \\
50  & {0.3969} & {0.4052} & {0.3459} & {0.4580} & {0.4183} & {0.3892} & {0.3836} & {0.3412} & {0.3333} & {0.4521} & {0.4118} & {0.3824} & {0.3704} & {0.3386} & {0.2807} & {0.4321} & {0.4021} & {0.3684} & {0.3942} & {0.3485} & {0.2897} & {0.4327} & {0.3900} & {0.3793} \\
100 & {0.6718} & {0.7712} & {0.8216} & {0.7023} & {0.8366} & {0.8216} & {0.6575} & {0.7882} & {0.8529} & {0.6849} & {0.8235} & {0.8529} & {0.6296} & {0.7831} & {0.8246} & {0.6790} & {0.8148} & {0.8333} & {0.4712} & {0.4564} & {0.4690} & {0.6538} & {0.7718} & {0.8069} \\
150 & {0.7023} & {0.8105} & {0.8216} & {0.7328} & {0.8366} & {0.8216} & {0.6986} & {0.7882} & {0.8529} & {0.7123} & {0.8235} & {0.8529} & {0.6914} & {0.8148} & {0.8333} & {0.7160} & {0.8254} & {0.8333} & {0.6442} & {0.7386} & {0.8345} & {0.7019} & {0.7967} & {0.8345} \\
\bottomrule
\end{tabular}
\end{adjustbox}
\caption{Performance comparison of EILOF and ILOF on the Shuttle dataset.}
\label{tab:performance_comparison_shuttle}
\end{table*}

\begin{table*}[ht]
\centering
\Large
\begin{adjustbox}{max width=\textwidth}
\begin{tabular}{c|ccc|ccc|ccc|ccc|ccc|ccc|ccc|ccc}
\toprule
\textbf{k (Neighbors)} & \multicolumn{6}{c|}{\textbf{m = 5}} & \multicolumn{6}{c|}{\textbf{m = 10}} & \multicolumn{6}{c|}{\textbf{m = 20}} & \multicolumn{6}{c}{\textbf{m = 40}} \\
 & \multicolumn{3}{c|}{\textbf{ILOF}} & \multicolumn{3}{c|}{\textbf{EILOF}} & \multicolumn{3}{c|}{\textbf{ILOF}} & \multicolumn{3}{c|}{\textbf{EILOF}} & \multicolumn{3}{c|}{\textbf{ILOF}} & \multicolumn{3}{c|}{\textbf{EILOF}} & \multicolumn{3}{c|}{\textbf{ILOF}} & \multicolumn{3}{c}{\textbf{EILOF}} \\
 & \textbf{5\%} & \textbf{7\%} & \textbf{10\%} & \textbf{5\%} & \textbf{7\%} & \textbf{10\%} & \textbf{5\%} & \textbf{7\%} & \textbf{10\%} & \textbf{5\%} & \textbf{7\%} & \textbf{10\%} & \textbf{5\%} & \textbf{7\%} & \textbf{10\%} & \textbf{5\%} & \textbf{7\%} & \textbf{10\%} & \textbf{5\%} & \textbf{7\%} & \textbf{10\%} & \textbf{5\%} & \textbf{7\%} & \textbf{10\%} \\
\midrule
50  & {0.2778} & {0.3902} & {0.4536} & {0.2778} & {0.3902} & {0.4536} & {0.2778} & {0.3902} & {0.4536} & {0.2778} & {0.3902} & {0.4536} & {0.2778} & {0.3879} & {0.4513} & {0.2778} & {0.3879} & {0.4513} & {0.2759} & {0.3855} & {0.4467} & {0.2759} & {0.3855} & {0.4467} \\
100 & {0.5139} & {0.6585} & {0.7732} & {0.5139} & {0.6585} & {0.7732} & {0.5139} & {0.6585} & {0.7732} & {0.5139} & {0.6585} & {0.7732} & {0.5139} & {0.6667} & {0.7795} & {0.5139} & {0.6667} & {0.7795} & {0.5241} & {0.6747} & {0.7716} & {0.5241} & {0.6747} & {0.7716} \\
150 & {0.6111} & {0.6951} & {0.7835} & {0.6111} & {0.6951} & {0.7835} & {0.6111} & {0.6951} & {0.7835} & {0.6111} & {0.6951} & {0.7835} & {0.6111} & {0.7030} & {0.7897} & {0.6111} & {0.6909} & {0.7897} & {0.6207} & {0.6988} & {0.7817} & {0.6207} & {0.6988} & {0.7817} \\
\midrule
\multicolumn{1}{c}{} & \multicolumn{6}{c|}{\textbf{m = 80}} & \multicolumn{6}{c|}{\textbf{m = 160}} & \multicolumn{6}{c|}{\textbf{m = 320}} & \multicolumn{6}{c}{\textbf{m = 640}} \\
 & \multicolumn{3}{c|}{\textbf{ILOF}} & \multicolumn{3}{c|}{\textbf{EILOF}} & \multicolumn{3}{c|}{\textbf{ILOF}} & \multicolumn{3}{c|}{\textbf{EILOF}} & \multicolumn{3}{c|}{\textbf{ILOF}} & \multicolumn{3}{c|}{\textbf{EILOF}} & \multicolumn{3}{c|}{\textbf{ILOF}} & \multicolumn{3}{c}{\textbf{EILOF}} \\
 & \textbf{5\%} & \textbf{7\%} & \textbf{10\%} & \textbf{5\%} & \textbf{7\%} & \textbf{10\%} & \textbf{5\%} & \textbf{7\%} & \textbf{10\%} & \textbf{5\%} & \textbf{7\%} & \textbf{10\%} & \textbf{5\%} & \textbf{7\%} & \textbf{10\%} & \textbf{5\%} & \textbf{7\%} & \textbf{10\%} & \textbf{5\%} & \textbf{7\%} & \textbf{10\%} & \textbf{5\%} & \textbf{7\%} & \textbf{10\%} \\
50  & {0.2484} & {0.3543} & {0.4251} & {0.2876} & {0.3886} & {0.4444} & {0.2500} & {0.3370} & {0.3945} & {0.2625} & {0.3696} & {0.4220} & {0.2722} & {0.3571} & {0.3915} & {0.2722} & {0.3571} & {0.3915} & {0.1867} & {0.2713} & {0.3062} & {0.1956} & {0.3101} & {0.3648} \\
100 & {0.4706} & {0.6286} & {0.7729} & {0.5229} & {0.6743} & {0.7826} & {0.4125} & {0.5652} & {0.7523} & {0.5000} & {0.6522} & {0.7706} & {0.4497} & {0.6327} & {0.7574} & {0.5444} & {0.6837} & {0.7489} & {0.3200} & {0.4031} & {0.3974} & {0.4356} & {0.5814} & {0.7427} \\
150 & {0.5882} & {0.6971} & {0.8019} & {0.5882} & {0.7086} & {0.7923} & {0.5750} & {0.6957} & {0.7890} & {0.5750} & {0.7065} & {0.7798} & {0.5917} & {0.7143} & {0.7660} & {0.6154} & {0.7143} & {0.7574} & {0.4800} & {0.6202} & {0.7818} & {0.5956} & {0.6977} & {0.7492} \\
\bottomrule
\end{tabular}
\end{adjustbox}
\caption{Performance comparison of EILOF and ILOF on the Credit Fraud dataset.}
\label{tab:performance_comparison_credit}
\end{table*}

Since the optimal threshold for the local outlier factor is unknown and the 7\% outlier proportion is not predetermined, we tested thresholds of 5\%, 7\%, and 10\% to evaluate performance. These thresholds represent the predicted outlier proportions determined by our algorithm, which may differ from the actual outlier proportion in the dataset. Additionally, we explored three different values of $k$ to examine how the size of the neighborhood influences the performance of the EILOF and ILOF algorithms.

As illustrated in Figure \ref{fig:three_graphs}, EILOF outperformed ILOF in terms of the F1 score. It is interesting to note that a 10\% threshold for both algorithms performs better than the 7\% and 5\% thresholds, despite the data containing 7\% outliers, when $k = 100$ and $k = 150$ (Figures \ref{fig:graph2} and \ref{fig:graph3}). However, for a smaller $k$ value ($k = 50$), the 5\% threshold yields better results than the other two thresholds (Figure \ref{fig:graph1}).

\subsection{Credit Card Fraud Dataset}

We further examined the effectiveness of the proposed EILOF algorithm using the Credit Card Fraud dataset from Kaggle \cite{dal_pozzolo_credit_card_fraud}, which is characterized by an extreme class imbalance: fraudulent transactions (outliers) constitute a very small fraction of the whole data. Specifically, fraud accounts for only 0.172\% of all transactions (492 frauds out of 284,807 transactions). Similar to the Shuttle dataset, our goal was to evaluate EILOF and ILOF in a streaming context under varying outlier thresholds (5\%, 7\%, and 10\%) and different neighborhood sizes ($k$).

To adapt the dataset to a streaming context, we constructed a reduced dataset in which outliers constituted approximately 5\% of the observations. Specifically, we first isolated all fraudulent (Class = 1) transactions and randomly sampled a subset of legitimate (Class = 0) transactions to reach the desired ratio of 5\% fraud. This choice was made to manage the high computational cost of ILOF when processing numerous streaming data points. The combined dataset was then sorted based on the original temporal order to preserve any time-dependent structure. Similar to the Shuttle dataset, the first 1,000 observations were used as the static baseline, while the subsequent 640 observations were considered as streaming data. Each subset was standardized independently to simulate real-world streaming conditions, in which incoming data may exhibit different statistical properties from those of the initial training set.

After partitioning and normalization, the EILOF and ILOF algorithms were progressively applied to detect fraudulent transactions in the streaming portion. Interestingly, while the overall dataset was initially designed to contain 5\% outliers, the 1,640 data points used in this experiment included approximately 7\% outliers. This discrepancy arose because the streaming subset was selected based on temporal order to preserve the real-world sequential nature of the data. Over time, the natural data distribution revealed localized clustering of outliers, reflecting real-world patterns where anomalies often concentrate in specific periods or regions due to shifts in behavior or fraud strategies. The initial reduction to 5\% outliers was done to reduce computational costs, especially for ILOF. However, maintaining a strict 5\% ratio in the streaming data would have altered the original sequence, making the data less realistic. By preserving the actual 7\% outlier proportion in the streaming subset, we ensured that the evaluation reflected real-world conditions more accurately. As in the Shuttle dataset experiments, $F_{1}$ scores were compared across thresholds of 5\%, 7\%, and 10\%, with the impact of varying neighborhood sizes ($k = 50, 100, 150$) also analyzed to evaluate detection performance.

\begin{figure}[!h]
    \centering
    \begin{subfigure}[b]{0.4\textwidth}
        \centering
        \includegraphics[width=\linewidth]{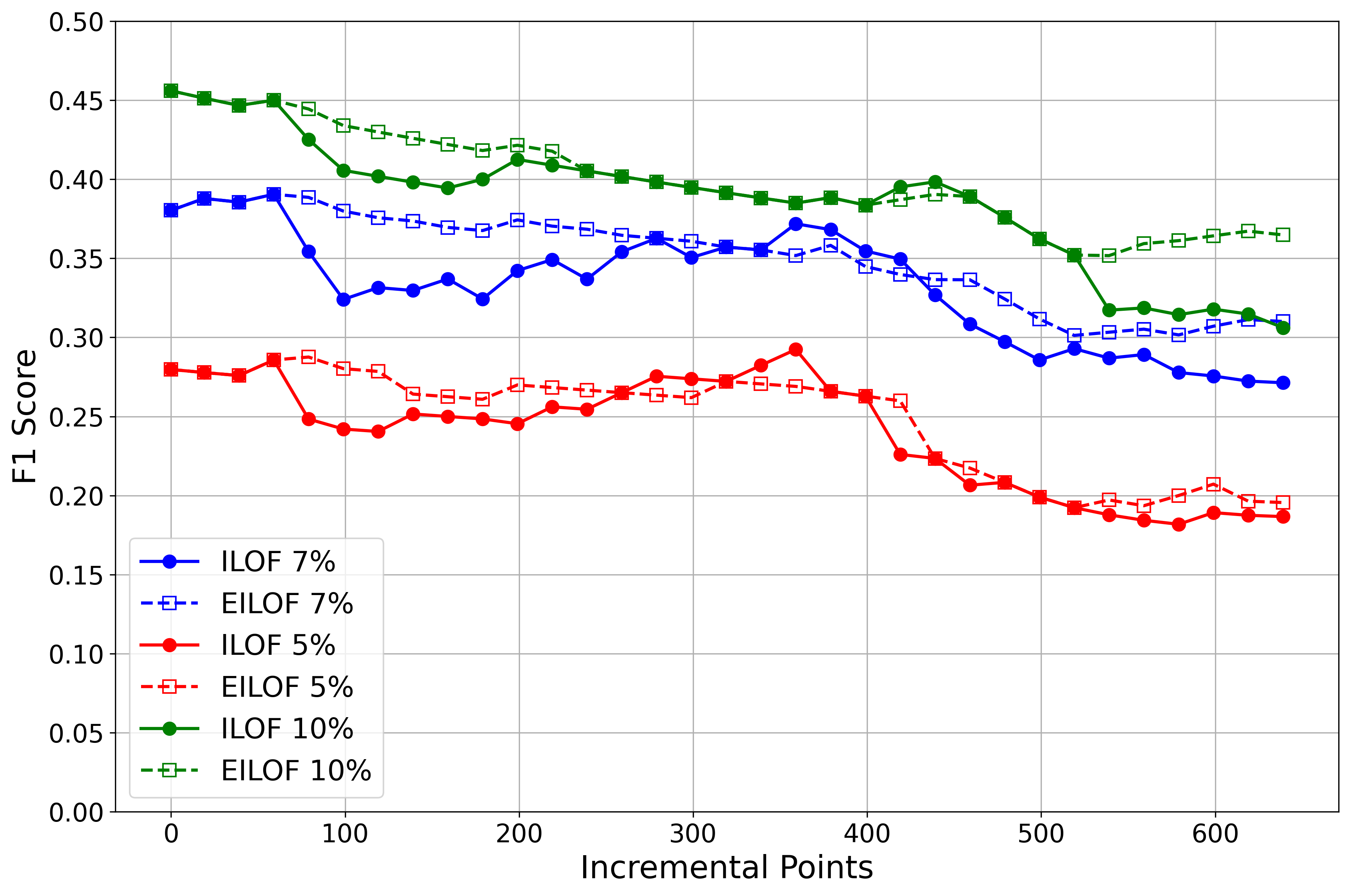}
        \caption{$k = 50$}
        \label{fig:graph4}
    \end{subfigure}
    \hfill
    \begin{subfigure}[b]{0.4\textwidth}
        \centering
        \includegraphics[width=\linewidth]{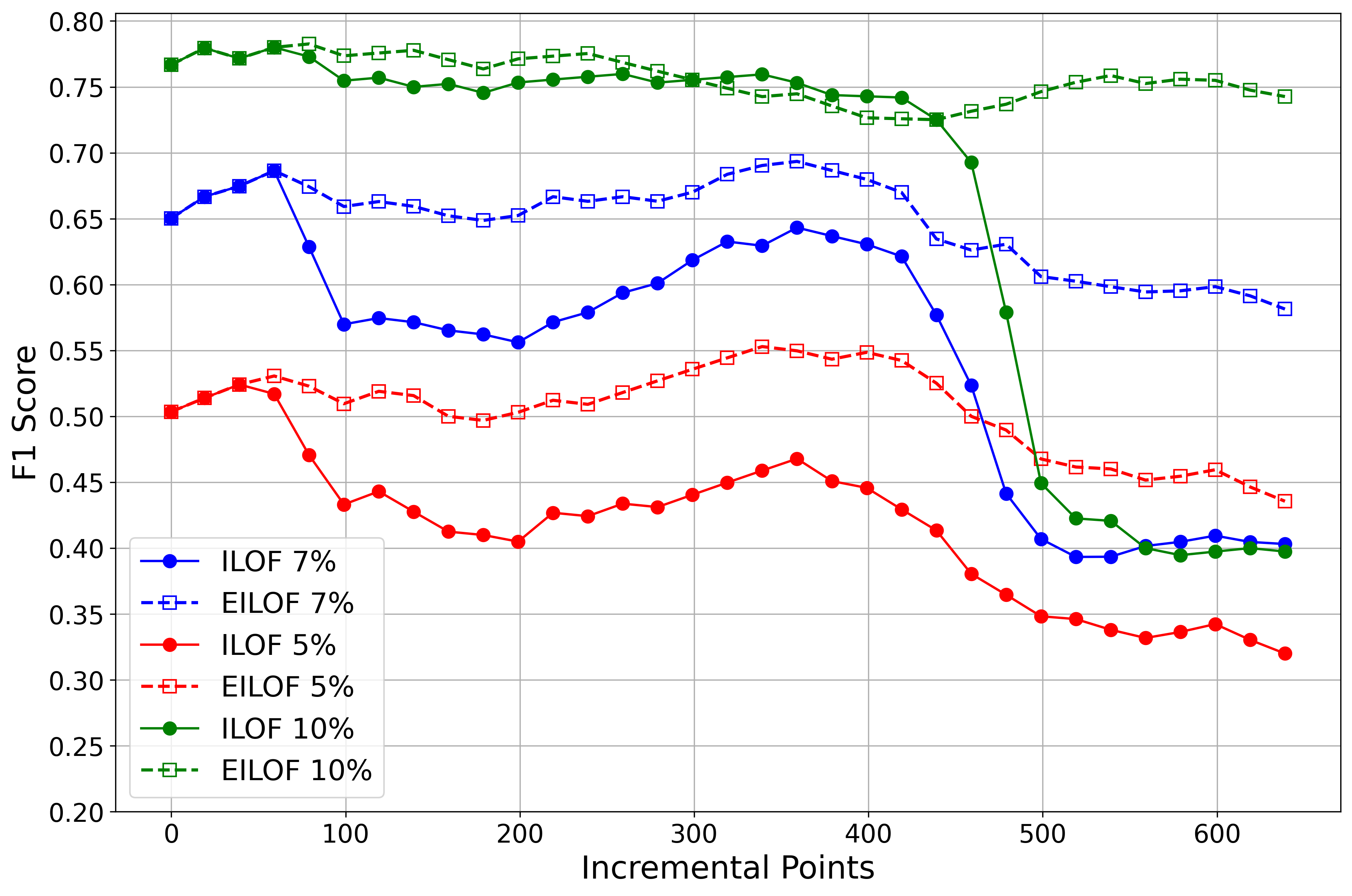} 
        \caption{$k = 100$}
        \label{fig:graph5}
    \end{subfigure}
    \hfill
    \begin{subfigure}[b]{0.4\textwidth}
        \centering
        \includegraphics[width=\linewidth]{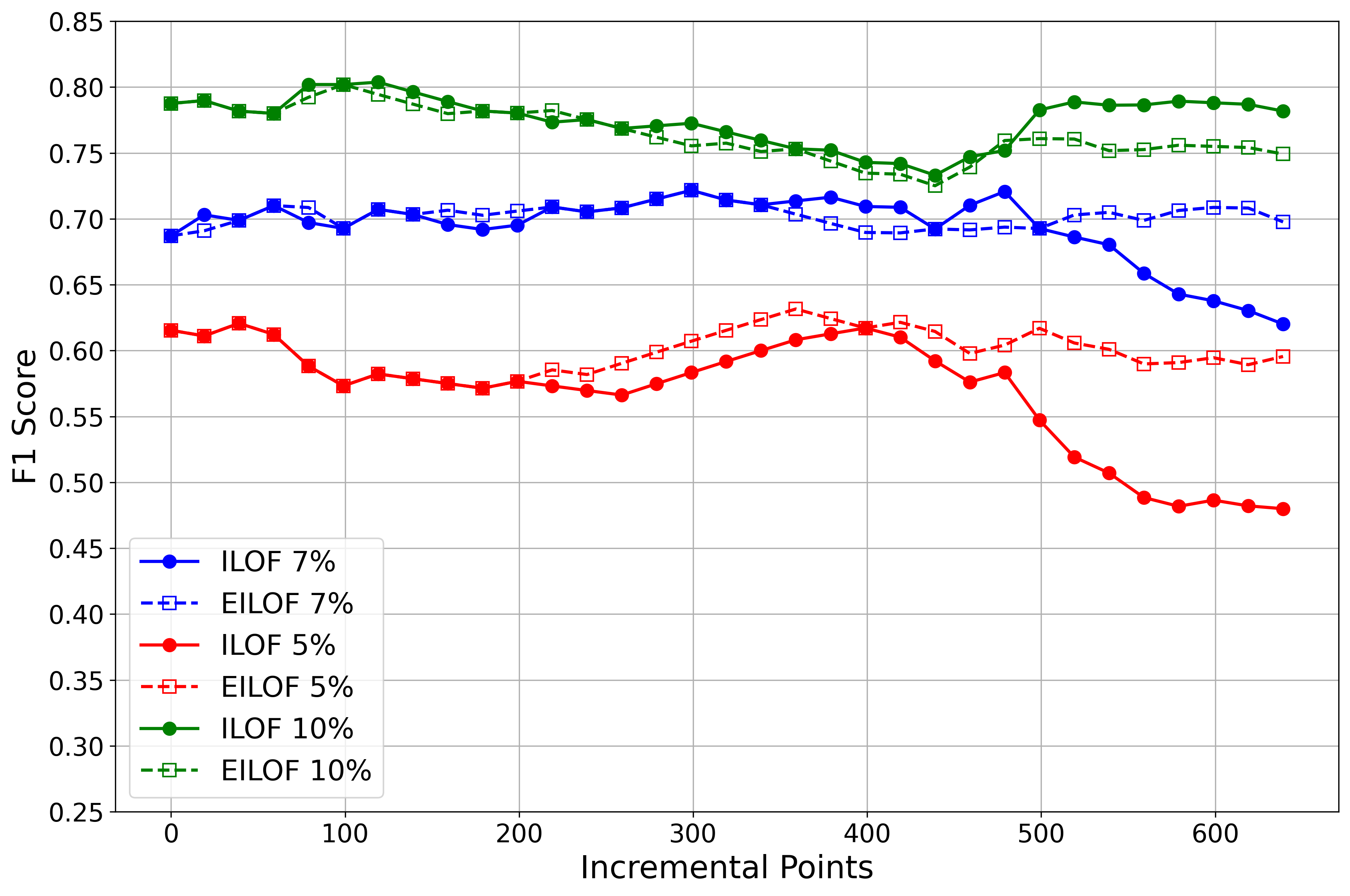} 
        \caption{$k = 150$}
        \label{fig:graph6}
    \end{subfigure}
    \caption{Comparison of EILOF and ILOF performance on Credit Card Fraud dataset. The $F_1$ scores are plotted against incremental points. Key points are selected every 20th increment to visualize the performance trend. (a) $k = 50$, (b) $k = 100$, (c) $k = 150$.}
    \label{fig:three_graphs_credit}
\end{figure}

As shown in Figure \ref{fig:three_graphs_credit}, EILOF consistently outperformed ILOF across most thresholds and $k$ values, similar to the observations in the Shuttle dataset. However, as illustrated in Figure \ref{fig:graph6}, when the incremental points exceed 480, the performance advantage of EILOF begins to decline and becomes less pronounced, particularly at higher $k$ values. This behavior could be attributed to the increased neighborhood size ($k$) reducing EILOF's sensitivity to subtle changes in the streaming data as the dataset becomes denser with additional incremental points. As discussed in Section~\ref{sec:ilof_performance}, larger neighborhood sizes provide more stable density estimates but may diminish the algorithm's ability to detect localized outliers.

The results across both datasets demonstrate that EILOF consistently outperforms ILOF across various thresholds and neighborhood sizes. Notably, at larger neighborhood sizes ($k = 100$ and $k = 150$), a 10\% threshold yielded the highest $F_1$ scores (Figures \ref{fig:graph2}, \ref{fig:graph3}, \ref{fig:graph5}, and \ref{fig:graph6}), even though the data for both subsets of the datasets contained a lower proportion of outliers (7\%).

This observation can be attributed to the trade-off between precision and recall, which significantly impacts $F_1$ scores, particularly in the context of varying $k$ values and threshold settings. These trade-offs significantly impact the algorithm’s ability to detect outliers, especially in dynamic datasets with varying densities and patterns.

A smaller $k$ value makes the algorithm more sensitive to local variations, improving recall because the algorithm can detect subtle outliers more effectively. However, this sensitivity can also lead to a higher number of false positives, where normal points are incorrectly identified as outliers, thereby reducing precision. This reduction in precision can be particularly problematic in highly dynamic datasets, where noise or fluctuating patterns are mistaken for anomalies. The ability to detect subtle outliers is advantageous in some cases but may compromise overall detection accuracy if the dataset exhibits rapid changes or transient outlier patterns. When the threshold is set to 5\%, even though the actual outlier proportion is 7\%, the algorithm’s sensitivity allows it to detect a significant portion of true outliers. The lower threshold reduces the number of points classified as outliers, which helps to reduce false positives and thus increases precision. The sacrifice in recall is minimal because the algorithm is already highly sensitive.

\section{Discussion}
In this section, we discuss the key factors that enable the EILOF algorithm to outperform the ILOF algorithm. A significant reason is that the EILOF algorithm does not alter the underlying LOF scores of the initial fixed dataset. Traditional online methods often struggle as more data streams in, especially because the fixed number of neighbors \( k \) becomes insufficient to accurately capture the characteristics of an evolving dataset.

Our approach addresses this issue by focusing on updating the LOF scores for newly added data points without recalculating the LOF scores for the existing data points. This updating strategy ensures that the historical data remains consistent and stable, preventing the fluctuations and inaccuracies that can occur with traditional methods. 

Specifically, as new data points are added, the algorithm incrementally updates the LOF scores only for these new points. By keeping the LOF scores of the original data points unchanged, we maintain the integrity of the initial analysis. This is crucial because changing the LOF scores of the initial data could lead to inconsistencies and reduced accuracy over time.

In contrast, if we were to continually recalculate the LOF scores for all data points each time new data is added, the fixed number of neighbors $k$  would no longer be sufficient. The neighborhood of each point would fail to accurately represent the local density, resulting in less precise outlier detection. However, this effect is mitigated in EILOF.

In summary, our method’s advantage lies in its ability to:
\begin{enumerate}
\item Preserve the integrity of the initial dataset by maintaining stable and consistent LOF scores.
\item Efficiently adapt to data streams by selectively updating LOF scores only for newly added points.
\item Minimize the limitations of a fixed number of neighbors \( k \), ensuring more accurate outlier detection as the dataset evolves.
\end{enumerate}

\section{Conclusions and Future Work}
In this paper, we propose an efficient LOF-based outlier detection algorithm, referred to as EILOF, for data streams. This method only computes the LOF scores of new data points without modifying the LOF scores of existing points. Due to the presence of noise inherent in most datasets, the performance of the EILOF algorithm is not degraded, but rather improved, by the slight loss in accuracy of the LOF computations in the whole updated dataset. This is because the marginal decrease in precision associated with the LOF calculations helps to mitigate the risk of overfitting. Numerical tests in both simulated environments and real-world data demonstrate that the EILOF algorithm outperforms the ILOF algorithm across various scenarios.

While the EILOF algorithm demonstrates significant improvements in efficiency and scalability, several areas for future work remain. Beyond adjusting \( k \) based on sample size, advanced optimization techniques for other parameters that may influence the algorithm’s performance, such as weighting schemes for neighbors in outlier detection, should be explored \cite{campos2016evaluation, schubert2014local}. For instance, the kNN-LOF algorithm, proposed in a recent work \cite{xu2022outlier}, assigns different weights to points based on their proximity to the point of interest, which can enhance EILOF’s outlier detection in datasets with varying density distributions .

Implementing a real-time version of the EILOF algorithm is another critical area, incorporating efficient data structures and parallel processing techniques to handle high-velocity data streams. This is crucial for applications requiring immediate anomaly detection. Since the performance of the algorithm also relies on whether data is outdated, it is worth exploring various techniques that prioritize recent data over outdated data, such as AMSD \cite{puttagunta2002adaptive} and CLOF \cite{yao2018incremental}. However, handling outdated data presents challenges because different industries face unique issues due to the varying lifecycles and relevance of data. For example, while financial data can become outdated quickly, medical data might retain its importance for longer periods. This variability necessitates adaptive strategies tailored to each industry’s needs. Exploring techniques that adjust to the changing relevance of data over time is necessary. Additionally, conducting extensive benchmarking against other state-of-the-art outlier detection methods across diverse and large-scale datasets is essential \cite{campos2016evaluation}. This comprehensive benchmarking will help validate the algorithm’s performance and robustness in various scenarios. Benchmarking across varied domains such as finance, healthcare, and cybersecurity will provide a comprehensive understanding of the algorithm’s robustness and generalizability \cite{hodge2004survey}.

By addressing these areas, we aim to further enhance the applicability and performance of the EILOF algorithm, making it a robust solution for real-time anomaly detection in various domains.

The codes and data for all examples are available from \href{https://pypi.org/project/eilof}{https://pypi.org/project/eilof}

\bibliography{bibliography}
\bibliographystyle{plain}

\end{document}